\documentclass[twocolumn]{aastex631}

\usepackage{array}
\usepackage{graphicx}
\usepackage{tabularx}
\usepackage{multirow}
\usepackage{amsmath}
\usepackage{esvect}
\begin{document}

\title{Comparative 3D Asymmetric Expansion and Angular Widths Evolution of Fast and Slow Coronal Mass Ejections}

\author[0009-0007-4956-5108]{Anjali Agarwal}
\affiliation{Indian Institute of Astrophysics, II Block, Koramangala, Bengaluru 560034, India}
\affiliation{Pondicherry University, R.V. Nagar, Kalapet 605014, Puducherry, India}

\author[0000-0003-2740-2280]{Wageesh Mishra}
\affiliation{Indian Institute of Astrophysics, II Block, Koramangala, Bengaluru 560034, India}
\affiliation{Pondicherry University, R.V. Nagar, Kalapet 605014, Puducherry, India}

\author[0000-0003-0951-2486]{Jie Zhang}
\affiliation{Department of Physics and Astronomy, George Mason University, Fairfax, VA, USA}

\author[0009-0006-3209-658X]{Soumyaranjan Khuntia}
\affiliation{Indian Institute of Astrophysics, II Block, Koramangala, Bengaluru 560034, India}

\author[0000-0003-4867-7558]{Manuela Temmer}
\affiliation{Institute of Physics, University of Graz, A-8010 Graz, Austria}

\correspondingauthor{Anjali Agarwal and Wageesh Mishra}
\email{anjaliagarwal1024@gmail.com; m.wageesh30@gmail.com}

\begin{abstract}

The radial and lateral dimensions of coronal mass ejections (CMEs) influence their duration and probability of encounter at Earth. These properties are linked to the expansion speed of CMEs in different radial and lateral directions; however, most earlier studies modeled CME evolution using a projected full ice-cream cone geometry, which does not distinguish between radial and lateral expansion. Our study investigates the asymmetric expansion (relative radial and lateral components) and kinematics of seven fast and seven slow CMEs within coronagraphic heights, using the Graduated Cylindrical Shell model. Our study confirms that CMEs expand asymmetrically, with lateral expansion exceeding radial expansion in both CME populations. This asymmetry limits the accuracy of the full ice-cream cone model. For both fast and slow CMEs, higher leading edge speeds are associated with higher expansion speeds. At a height of 10 $R_\odot$, slow CMEs with larger expansion speeds (lateral and radial) have larger angular widths (face-on and edge-on), whereas fast CMEs exhibit a negative correlation between lateral expansion speed and face-on angular width. We find that the expansion and propagation speeds of slow CMEs exhibit a two-phase evolution, whereas those of fast CMEs display more diverse trends. Overall, this study suggests that fast and slow CMEs evolve differently and should not be treated as a single population in statistical estimates of their physical parameters. Our study highlights the importance of estimating CME angular widths and expansion speeds along different directions, and beyond standard coronagraphic heights, to capture their complete physical evolution.

\end{abstract}


\section{Introduction} \label{sec:int}

Coronal mass ejections (CMEs) are large-scale eruptions of magnetized plasma from the Sun into the heliosphere and are primary drivers of space-weather events \citep{Schwenn2005,Pulkkinen2007,Webb2012,Schrijver2015,Zhang2021,Mishra2023,Temmer2024}. Accurate prediction of their arrival at Earth requires reliable estimates of CME expansion speeds and angular widths \citep{Yashiro2004,Gopalswamy2009,Howard2012,Mostl2014,Scolini2020,Mishra2023,Agarwal2026}. These properties can be inferred from remote and in situ observations, particularly from multiple spacecraft, but each approach has inherent limitations. Owing to their optically thin and three-dimensional nature, remote observations require correction for projection effects \citep{Sheeley1999,Xie2004,Thernisien2006,Liu2010,Davies2013,Mishra2014a,Amerstorfer2018,Agarwal2024,Nikou2025}, while in situ measurements are subject to geometric selection effects that limit inference of CME global structure \citep{Burlaga1981,Gosling1990,Crooker1996,Bothmer1998,Zurbuchen2006,Mishra2023}.

Combined analyses of observations and modeling suggest that CME dynamics are dominated by the Lorentz force near the Sun and by viscous interactions with the solar wind and large-scale heliospheric structures in interplanetary space \citep{Subramanian2007,Vrsanak2007,Temmer2011,Sachdeva2015,Mishra2013,Mishra2014,Mishra2015a,Mishra2017}. CME expansion along different spatial (radial or lateral) dimensions is caused by higher internal pressure than the ambient medium and plays a crucial role in shaping their three-dimensional kinematics, morphology, size, arrival time, and probability of Earth encounter \citep{Dasso2006,Gulisano2010,Kilpua2017,Mishra2021a,Agarwal2024}. There have been numerous earlier studies on the evolution of propagation speed; however, only limited studies have examined the expansion speed of CMEs utilizing their 3D characteristics (e.g., \citealt{Savani2009,Zhuang2023,Agarwal2024}).

CME expansion along different spatial dimensions plays a crucial role in shaping their three-dimensional kinematics, morphology, size, arrival time, and probability of Earth encounter \citep{Kilpua2017}. Earlier studies have estimated the lateral (perpendicular to the CME's propagation direction) expansion speed ($V_{{exp}}$) of limb CMEs to empirically establish its relation with radial propagation speed ($V_{{rad}}$) (e.g., \citealt{Dallago2003,Schwenn2005,Gopalswamy2009,Gopalswamy2010,Gopalswamy2012}). The lateral expansion speed of a CME is estimated using the change in the full base length (cone's full angular width) of the assumed cone geometry. From these studies, the widely used empirical relation is $V_{{rad}} = 0.88 V_{{exp}}$ \citep{Dallago2003,Schwenn2005} which can be translated to $V_{{rad}} = 1.76 V_{{exp}}$, if lateral expansion is estimated measuring the change in the half base length (cone's half angular width) of the assumed cone geometry. Such empirical relations have often been utilized to derive the radial propagation speed of Earth-directed halo CMEs \citep{Dallago2004,Gopalswamy2012,Makela2016,Scolini2019,Mishra2021a}.

Expanding on previous works of \citet{Zhao2002} and \citet{Xie2004}, three mathematical functions (assuming flat cone, shallow, and full ice cream cone models) are established by \citet{Gopalswamy2009} to connect the radial propagation speed (POS projected), lateral expansion speed, and half angular width of hundreds of limb CMEs. A subsequent study of \citet{Michalek2009} analyzed hundreds of CMEs and found $V_{{rad}} = 2.34 V_{{exp}}$ (corresponding to the cone's half base length) that match best, consistent with the findings from the full ice cream cone model of \citet{Gopalswamy2009}. For halo CMEs, it is suggested that their deprojected radial speed can be estimated using their angular width (based on the width-radial speed relationship for limb CMEs), longitude, and POS projected speed (e.g., \citealt{Gopalswamy2009,Gopalswamy2010}).

In studies employing empirical relations and derived mathematical functions (e.g., \citealt{Dallago2003,Schwenn2005,Gopalswamy2009,Michalek2009,Makela2016,Balmaceda2020,Mishra2021a}), a major source of uncertainty arises from the assumption of conical CME geometry. Throughout this study, the \textit{plane of CME propagation} is defined as the plane passing through the Sun center and containing both the CME propagation direction and the flux-rope axis. The \textit{plane perpendicular to the CME propagation} is defined as the plane passing through the flux-rope center and normal to the CME propagation direction. These two mutually orthogonal planes would define three dimensions of CME along the three mutually perpendicular directions: two dimensions in the plane of CME propagation and one dimension in the plane perpendicular to the CME propagation. Thus, in conical geometries adopted in earlier studies (e.g., full ice-cream cone model), there also exist two lateral (perpendicular to CME propagation direction) dimensions of CMEs; however, both (one lying in the plane of CME propagation and the other in the plane perpendicular to the CME propagation) dimensions are assumed to be equal \citep{Gopalswamy2009,Gopalswamy2012}. Consequently, lateral expansion speed is not distinguishable along these two lateral directions. Moreover, the lateral expansion speed is typically assumed to be identical to the radial expansion speed along the propagation direction of CME.

Physically, flux ropes associated with CMEs typically originate with two legs anchored to the Sun \citep{Chen1993,Michels1997,Chen2000}. In white-light coronagraph observations, however, the CME is primarily identified through its leading edge and flanks rather than its underlying magnetic footpoints. Therefore, approximating such a three-dimensional structure with a projected cone can yield inaccurate estimates of angular width that depend on the flux rope's orientation (tilt) relative to the observer, even at fixed latitude and longitude. Moreover, many earlier studies estimate constant speeds and angular widths averaged over the observed interval at larger coronal heights, and empirical relations derived for limb CMEs may still suffer from projection effects.

Moreover, in 3D, a CME evolves along three mutually perpendicular directions, and the expansion speeds along these three directions need not be the same \citep{Cremades2020}. Accordingly, the radial size (flux rope's cross-sectional extent along the CME's propagation direction) of a CME is governed by its radial expansion, while its lateral extent (in the plane of the CME propagation direction) is controlled by lateral expansion in the interplanetary (IP) medium. Despite this inherently three-dimensional nature, most existing studies of CME expansion rely on plane-of-the-sky (POS) projected coronagraphic observations \citep{Balmaceda2020}. To minimize projection effects, these studies typically focus on close-to-limb CMEs and employ different projected cone models with constant angular widths (e.g., \citealt{Dallago2003,Schwenn2006,Gopalswamy2009, Michalek2009,Mishra2021a}). Earlier approaches of adopting cone geometries cannot fully capture dimension-dependent CME expansion, particularly the lateral and radial components discussed above. Building on previous studies, a logical next step is to employ 3D reconstruction techniques to more accurately estimate the angular width and expansion speeds (along and perpendicular to the CME propagation direction) of different types of CMEs.

Despite the limitations of these empirical relations based on cone geometries, they are widely used to study CME expansion near the Sun and its solar cycle dependence (e.g., \citealt{Vourlidas2017,Balmaceda2020,Mishra2021a}), as they provide a simple means to estimate key CME parameters without complex 3D reconstruction. It is essential to apply 3D reconstruction techniques to CMEs, such as the Graduated Cylindrical Shell (GCS) model \citep{Thernisien2006,Thernisien2011}, and derive their 3D parameters to overcome the limitations of assuming a CME as a cone. The GCS model has been widely used in the literature to determine the de-projected (3D) kinematics of CMEs (e.g., \citealt{Mishra2015,Agarwal2024,Khuntia2025,Kay2024,Dupertuis2025}).

The GCS model approximates a CME as a hollow croissant structure consisting of two conical legs attached to a torus-shaped curved front. This geometry allows estimation of the CME spatial dimensions along different directions, as well as the angular widths corresponding to the GCS structure viewed in face-on and edge-on orientations. Therefore, adopting GCS geometry for CMEs can provide the estimates of their lateral and radial extents and expansion speeds \citep{Cabello2016}. The model assumes that the radial extent in the CME propagation plane is equal to perpendicular to the flux rope axis in the perpendicular plane (as discussed in Appendix~\ref{sec:appendix}). Consequently, the radial expansion speed along the CME propagation direction is equal to the lateral expansion speed in the perpendicular plane. However, this framework enables us to investigate the differences between the radial expansion and the lateral expansion along the flux rope axis within the propagation plane. Moreover, the more reliable estimates of derived 3D parameters of the GCS model require observations of the same CME from at least two viewpoints with sufficient angular separation \citep{Verbeke2023}.

We emphasize that only a limited number of studies have used the 3D hollow-croissant geometry adopted in the GCS model \citep{Thernisien2011} to estimate the evolution of CME angular widths and expansion speeds in different directions  \citep{Cabello2016,Cremades2020}. These studies were based on either a case study or a small number of CMEs, thus limiting the statistical validity. Also, these studies did not compare their estimates with the earlier-established empirical relations and mathematical functions derived from a geometrical cone model. Given the vast amount of literature that employs empirical relations, we believe that further research is needed on this subject, utilizing estimates of CMEs' 3D parameters.

The present study aims to enhance our understanding of the lateral and radial expansion speeds of CMEs by using 3D estimates of CME geometrical parameters within the GCS model framework. We are motivated to examine the asymmetry in lateral and radial expansion speeds and compare them with estimates from cone models widely adopted in earlier studies. In the GCS model framework, the GCS geometry viewed face-on and edge-on angular widths are basically along the two lateral dimensions of the CME. The face-on angular width is measured along the toroidal direction (along the axis of the torus), whereas the edge-on angular width is along the poloidal direction (lying in planes orthogonal to the toroidal axis).

Unlike previous studies (e.g., \citealt{Cabello2016,Cremades2020}), we investigate whether variations in angular width and asymmetric expansion are preferentially associated with fast or slow CMEs based on their deprojected 3D speeds. To further constrain CME evolution in the interplanetary medium, we examine the relationships between CME propagation speed (leading edge (LE)), expansion speeds (radial and lateral), and angular widths in different directions.

The paper is organized as follows: Section~\ref{sec:obs3D} details the CME selection, classification into fast and slow types, and analysis of their 3D kinematics, radial and lateral expansion, and their comparison with the estimates of expansion from the full ice-cream cone model (hereafter referred to as the full cone model throughout this manuscript). Section~\ref{sec:linrel} explores the correlations of CME LE speeds with face-on angular width along the toroidal direction and edge-on along the poloidal direction, propagation speeds of CME with radial and lateral expansion speed. This section also examines the relationships between lateral expansion speed and face-on angular width, and between radial expansion speed and edge-on angular width. Section~\ref{sec:resdis} presents the results, discussion, and implications for future work. Section~\ref{sec:conclu} presents the conclusion and novelty of the study.

\section{Implementing GCS Model for Deriving 3D Kinematics, Angular Widths, and Expansion Speeds of Selected CMEs}{\label{sec:obs3D}}

In this study, we investigate the evolution of de-projected angular widths and lateral and radial expansion speeds for 14 selected CMEs. We analyze the 3D kinematics of selected CMEs using the Graduated Cylindrical Shell (GCS) model \citep{Thernisien2009}. We implement the GCS model to the selected events utilizing simultaneous coronagraph observations from the Large Angle and Spectrometric COronagraph (LASCO) onboard \textit{Solar and Heliospheric Observatory} (\textit{SOHO}), which provides a field of view (FOV) of $\sim$2.2--30 $R_\odot$ through its C2 and C3 coronagraphs \citep{Brueckner1995}, together with the COR1 (FOV: 1.5--4 $R_\odot$) and COR2 (FOV: 2.5--15 $R_\odot$) instruments onboard the twin \textit{Solar TErrestrial RElations Observatory} (\textit{STEREO}) spacecraft \citep{Howard2008,Kaiser2008}. The combined multi-viewpoint observations enable the determination of more reliable 3D CME parameters \citep{Verbeke2023}.

The events are selected based on the following criteria: (i) they occur during the \textit{STEREO} era from late 2009 to early 2014, when the angular separation between \textit{SOHO} and the twin \textit{STEREO} spacecraft ranges from $60^\circ$ to $150^\circ$ to determine the more reliable 3D parameters \citep{Nikou2025}; and (ii) they are isolated in coronagraphic images up to their tracked heights, ensuring that their kinematics and morphology are not influenced by following CMEs. The GCS model represents a CME flux rope as a hollow croissant-shaped structure (two conical legs attached to a torus-shaped curved front) and includes six free parameters: latitude ($\theta$) and longitude ($\phi$), the half angle between the two conical legs ($\alpha$), the tilt angle ($\gamma$), the aspect ratio ($\kappa$), and the height of the leading edge ($h_f$). For completeness, a brief description of the GCS model is provided in Appendix~\ref{sec:appendix}, where we revisit the spatial and angular dimensions of the hollow croissant-shaped CME structure and relate them to the radial and lateral expansion speeds. In this study, we first apply the GCS model to contemporaneous coronagraph observations from STEREO/COR2 and SOHO/LASCO-C2, where the CME has sufficiently developed to be identified as a hollow croissant-shaped structure. Although GCS forward modeling is not unique and different parameter combinations can yield comparable fits to the observations, these viewpoints provide reasonable constraints for obtaining a representative set of GCS parameters for subsequent analysis. These parameters are then used as initial constraints for fitting the earlier STEREO/COR1 observations at lower heights.

\begin{figure*}
    \centering
    \includegraphics[scale= 0.4,trim={0cm 0cm 0cm 0cm},clip]{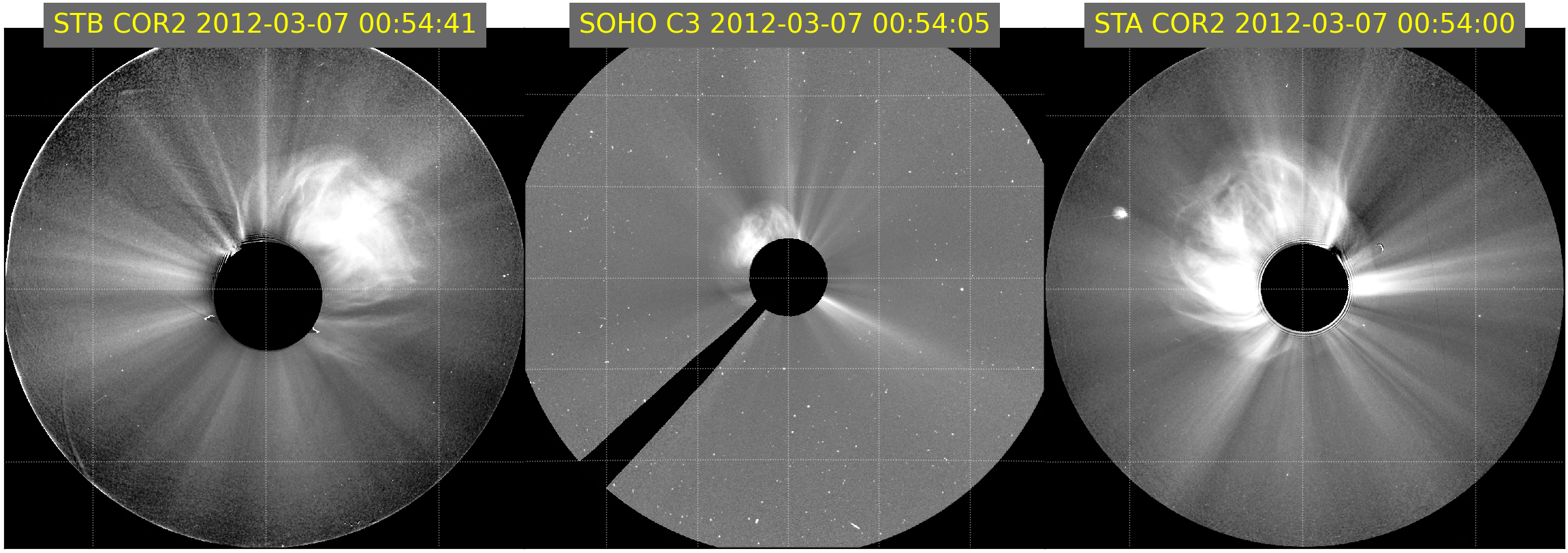}
    \includegraphics[scale= 0.4,trim={0cm 0cm 0cm 0cm},clip]{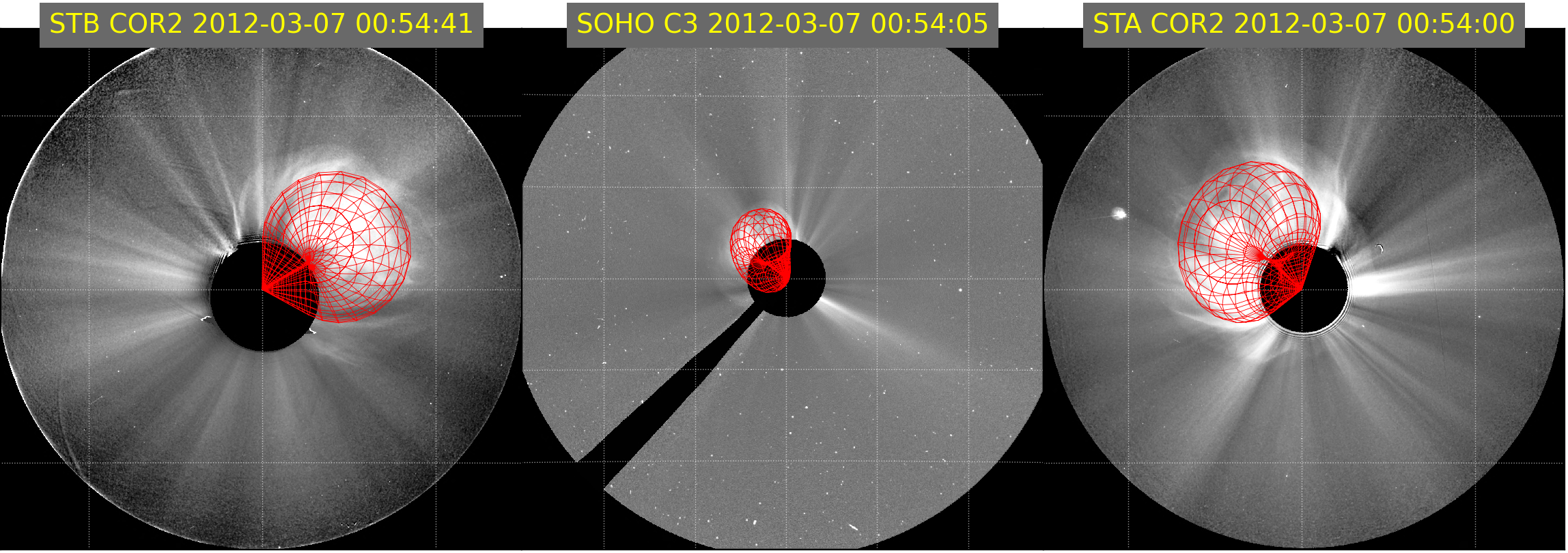}
    \caption{The top and bottom panels show the coronagraph observations of a fast CME without and with GCS fitting, respectively. The bottom panel shows the GCS wireframe overlaid in red on the observed CME. The fitting is performed using simultaneous coronagraph observations from three viewpoints: STEREO-B/COR2 (left), SOHO/LASCO C2 (center), and STEREO-A/COR2 (right). STB and STA denote the STEREO-B and STEREO-A spacecraft, respectively.}
    \label{fig:gcs_fit_fast}
\end{figure*}

\begin{figure*}
    \centering
    \includegraphics[scale= 0.4,trim={0cm 0cm 0cm 0cm},clip]{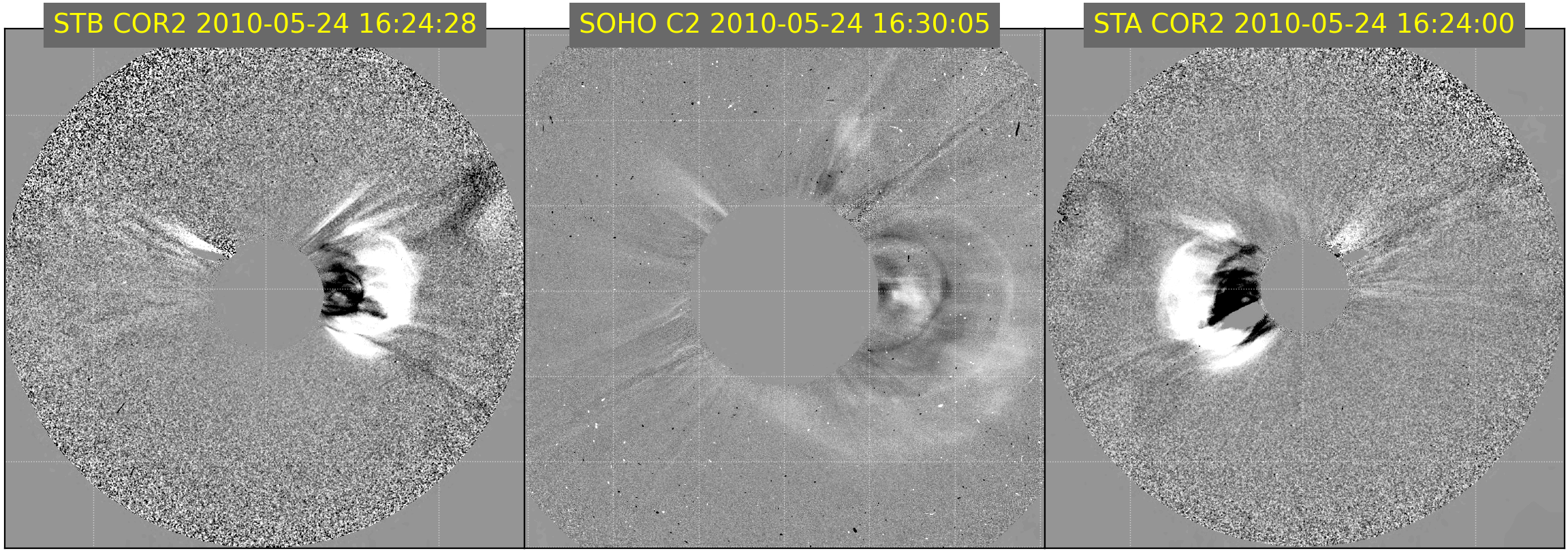}
    \includegraphics[scale= 0.4,trim={0cm 0cm 0cm 0cm},clip]{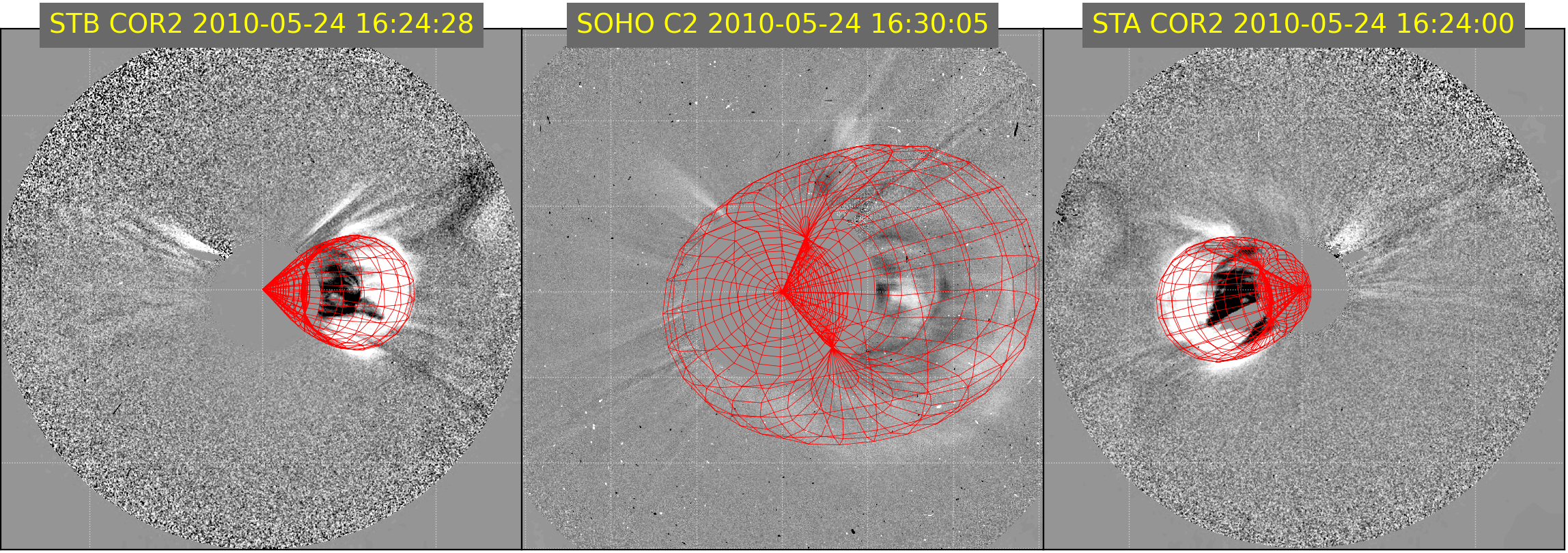}
    \caption{The top and bottom panels show the coronagraph observations of a slow CME without and with GCS fitting, respectively. The bottom panel shows the GCS wireframe overlaid in red on the observed CME. The fitting is performed using simultaneous coronagraph observations from three viewpoints: STEREO-B/COR2 (left), SOHO/LASCO C2 (center), and STEREO-A/COR2 (right). STB and STA denote the STEREO-B and STEREO-A spacecraft, respectively.}
    \label{fig:gcs_fit_slow}
\end{figure*}

\begin{table}
    \centering
    \small
    \begin{tabular}{ccccccc}
    \hline
    Date & $\theta~(^\circ)$ & $\phi~(^\circ)$ & $h_f$ ($R_\odot$) & $\alpha~(^\circ)$ &  $\kappa$ & $\gamma~(^\circ)$  \\
    \hline
    2009 Dec 16 & 19 & 0 & 14.3 & 11 & 0.17 & 5.04 \\
    2010 Feb 07 & -10 & 340 & 23.2 & 12  & 0.57 & -35 \\
    2010 Apr 03 & -24 & 3 & 13.7 & 25.5 & 0.37 & 9.79 \\
    2010 Apr 08 & -5 & 355 & 13.8 & 50 & 0.28 & -30 \\
    2010 May 24 & -2 & 20 & 14.1 & 25 & 0.6 & 15 \\
    2010 Jun 16 & 3 & 350 & 17.5 & 20 & 0.21 & -40 \\
    2011 Feb 01 & 10 & 190 & 13.3 & 20 & 0.45 & 75 \\
    2011 Mar 07 & 30 & 40 & 20.7 & 30 & 0.44 & -80 \\
    2011 Aug 04 & 20 & 30 & 17.8 & 42 & 0.30 & -60 \\
    2011 Sep 24 & 17 & 315 & 17.7 & 30 & 0.62 & -60 \\
    2012 Mar 07 & 30 & 330 & 10 & 30 & 0.5 & -75 \\
    2012 Jul 12 & -14 & 6 & 15 & 25 & 0.61 & 40 \\
    2012 Sep 28 & 20 & 20 & 17.3 & 40 & 0.55 & -75 \\
    2014 Jan 07 & -27 & 35 & 18 & 25 & 0.58 & 20 \\
    \hline 
    \end{tabular}
    \caption{The table lists the GCS parameters of the CME at the last tracked height. The columns, from left to right, represent the date, latitude ($\theta$), longitude ($\phi$), last tracked height ($h_f$), half angle ($\alpha$), aspect ratio ($\kappa$), and tilt ($\gamma$) of the CME.}
    \label{tab:tab_1}
\end{table}

The launch dates of the 14 selected CMEs, along with their 3D parameters derived from the GCS model at the final height of the tracked CME LE, are listed in Table~\ref{tab:tab_1}. We note that, for all selected events, the angular separation between the CME propagation direction and either of the twin STEREO spacecraft is $\gtrsim 45^\circ$. As a result, at least one STEREO spacecraft observes the CME closer to the limb, reducing the influence of projection effects relative to the Earth-based perspective \citep{Shen2013}. The only exception is the CME of 2014 Jan 07, for which the separation between the de-projected CME propagation direction and STEREO-B is $\sim5^\circ$ (halo for STEREO-B), whereas the corresponding separation with STEREO-A is $\sim65^\circ$.

We note that the GCS model has been implemented for a subset of our selected events in earlier studies \citep{Temmer2021,Kay2024}. However, the GCS parameters derived in this study may differ from previously reported values because the visual identification of CME boundaries and manual adjustment of model parameters introduce user-dependent variability. Consequently, uncertainties in the derived GCS parameters are inevitable. To quantify such uncertainties, \citet{Thernisien2009} performed a sensitivity analysis and reported mean uncertainties of approximately $\pm1.8^\circ$ in latitude, $\pm4.3^\circ$ in longitude, $\pm22^\circ$ in tilt angle, $^{+13^\circ}_{-7^\circ}$ in $\alpha$, $\pm0.48~R_\odot$ in height, and $^{+0.07}_{-0.04}$ in aspect ratio. Using an independent approach in which multiple researchers fitted the same CME events separately, \citet{Pluta2019} reported uncertainties of about $\pm5^\circ$ in latitude, $\pm5^\circ$ in longitude, $\pm30^\circ$ in tilt angle, $\pm10^\circ$ in $\alpha$, $\pm0.5~R_\odot$ in height, and $\pm0.025$ in aspect ratio. Both studies estimated these uncertainties for CMEs at heliocentric distances of around $10~R_{\odot}$. Consistent with earlier estimates of uncertainties, we adopt uncertainties of 5\% in height and 0.05 in aspect ratio. For the parameter $\alpha$, we assume an uncertainty of $10^\circ$ for all CMEs, except for the 2009 December 16 and 2010 February 07 events. Since these CMEs have $\alpha \sim 10^\circ$, we adopt a smaller uncertainty of $5^\circ$ in $\alpha$, corresponding to 50\% of its value. Except for the height uncertainty, which scales with CME height, the uncertainties assigned to the remaining GCS parameters are assumed to remain constant over the tracked CME height range.

All the selected CMEs are isolated in the coronagraphic images. We note, however, that the 2012 March 7 CME interacted with a following CME launched approximately one hour later and has been widely studied \citep{Patsourakos2016,Soni2023}. In this study, we therefore restrict our analysis to the kinematic evolution of this CME prior to the interaction. We estimate the LE speed of selected CMEs by utilizing the moving box linear fit technique on the observed height-time measurements of the CME LE (as described in \cite{Agarwal2024}). Figures~\ref{fig:gcs_fit_fast} and \ref{fig:gcs_fit_slow} show representative fast and slow CMEs, respectively. For each figure, the top panel displays the CME without GCS fitting, while the bottom panel shows the corresponding GCS reconstruction in the Stonyhurst heliographic coordinate system. The GCS fitting for all other CMEs is included as supplementary material. In the following section, we describe the criteria used to classify the selected CMEs as fast or slow, and present their kinematics.

\subsection{3D kinematics of selected CMEs}{\label{sec:3Dkin}}

The classification of CMEs into different speed categories is subjective, as it depends on selected height ranges, as some CMEs show significant dynamic evolution from lower to upper corona. The classification based on the speeds towards end data points may not be appropriate, as CME fronts are poorly defined in coronagraph images at higher heights, where the CME may become diffuse, increasing measurement uncertainties. On the other hand, classification based on the speed at the first data point may not be robust and obscure the dynamic evolution. Earlier studies have shown that CME acceleration typically peaks at low coronal heights ($<3~R_\odot$) \citep{Gallagher2003}. Therefore, we classify CMEs as fast or slow based on their average LE speeds estimated up to $3~R_\odot$. For events whose first tracked LE height is available only above $3~R_\odot$, the average speed is instead estimated up to $5~R_\odot$. This criterion ensures that the speeds used for classification are consistently derived within the middle corona ($\sim1.5$--$5~R_\odot$), where CME dynamics are still primarily governed by their internal magnetic forces \citep{West2023}. Classifying CMEs by their speeds in the middle corona enables a more direct comparison of their expansion speeds, which are primarily governed by their internal magnetic energy. At these heights, CMEs are less affected by overlying structures in the inner corona and by later interactions with the solar wind in the outer corona.

CMEs with average speeds at the first two measured data points exceeding 600 km s$^{-1}$ are classified as fast, while those with average speeds below this threshold are classified as slow. The adopted threshold is significantly higher than the typical fast solar wind and local magnetosonic speeds near the Sun \citep{Warmuth2005,Zucca2014}. Previous statistical studies, often focused on projected speeds, have also employed similar thresholds when identifying dynamically distinct CMEs \citep{Pant2021,Kumari2023}. To maintain a conservative and unambiguous classification, CMEs with slow to moderate speeds (below 600 km s$^{-1}$) were grouped under the slow CME category, while only those clearly exceeding this threshold were considered fast CMEs. This criterion ensures that the CMEs categorized as fast exhibit significant dynamical contrast with the background solar wind. Moreover, fast CMEs can drive shock waves, and tracking the shock front rather than the flux-rope structure can introduce significant uncertainties in the derived 3D parameters. However, these uncertainties can be reduced by tracking the front of the CME flux rope in base-difference images instead of running-difference images \citep{Hess2014}. Accordingly, in this study, we used base difference images to track fast CMEs and took extra care during GCS fitting to exclude the shock front, thereby enabling a reliable determination of CME parameters.

\begin{figure*}
\centering
\includegraphics[scale=0.73,trim={0.3cm 0.25cm 0cm 0.24cm},clip]{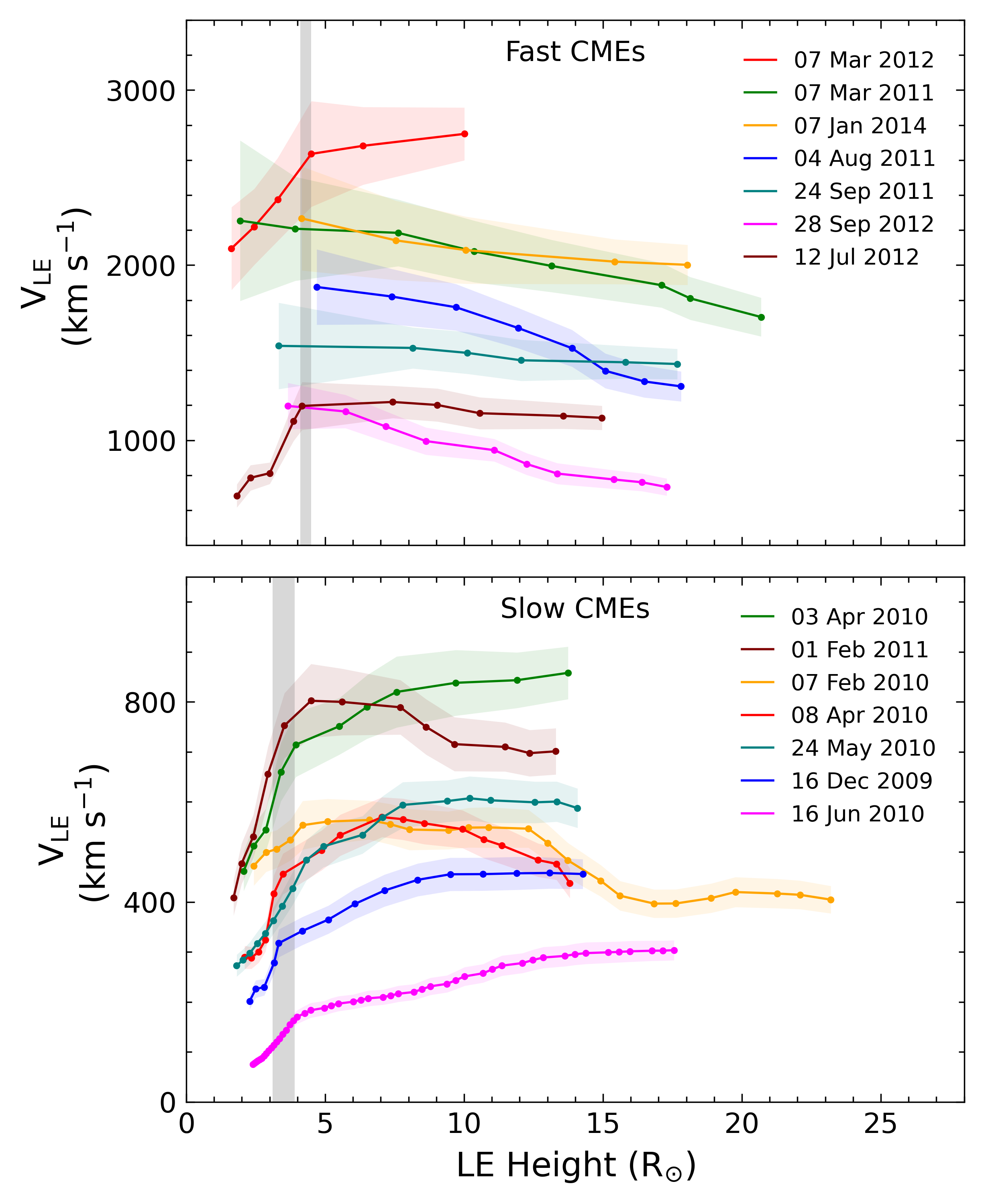}
\includegraphics[scale= 0.73,trim={0.2cm 0.25cm 0.2cm 0.24cm},clip]{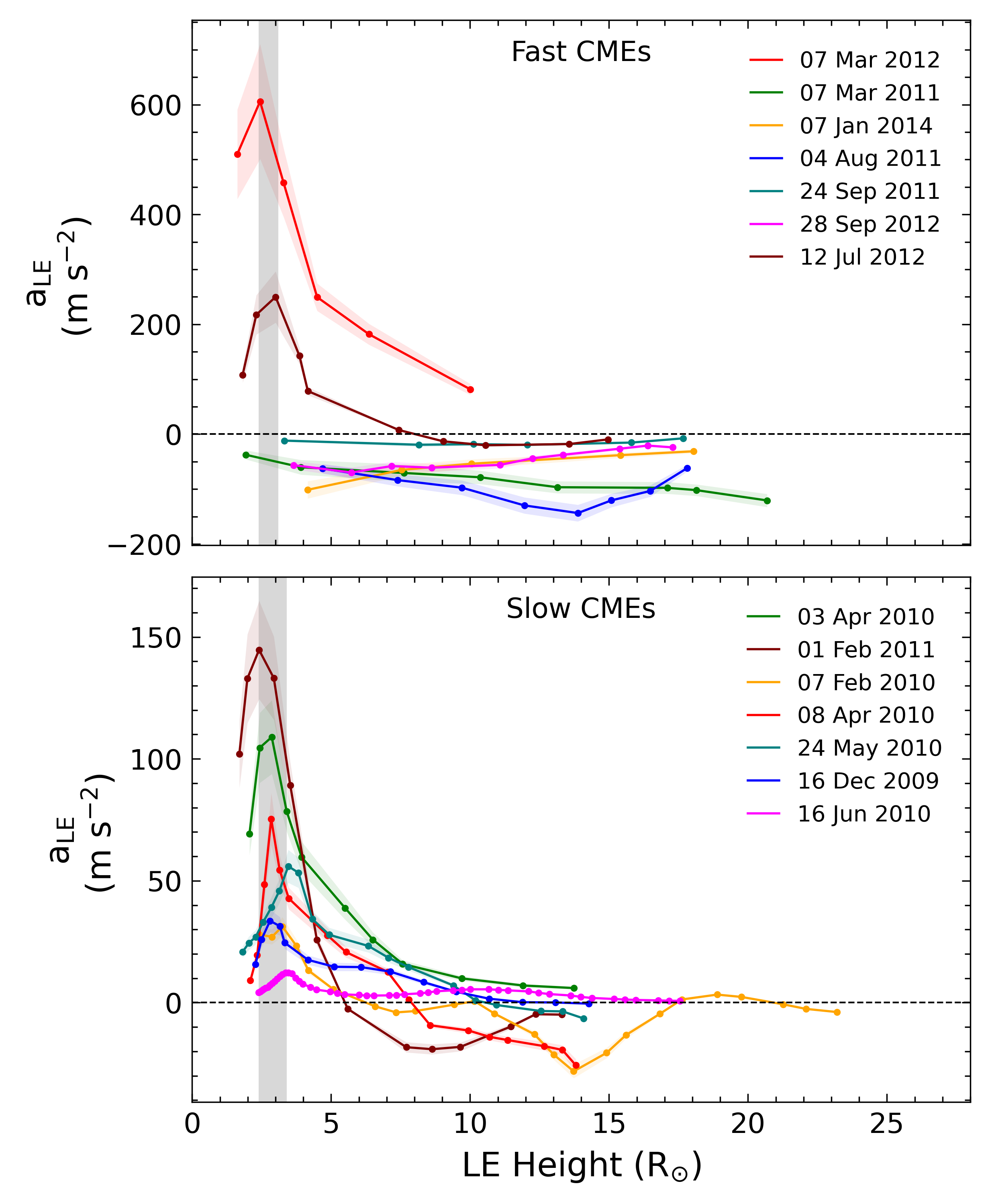}
\caption{The left and right columns represent the LE speed and acceleration, respectively, with the height of the CME LE, for fast CMEs in the top panel and for slow CMEs in the bottom panel. The legends in both panels are arranged in decreasing order of CME LE speed (from highest to lowest). The error bars are represented by transparent shaded regions in the same colors as the corresponding data points. In the left panel, the grey-shaded region spans the height range where the constant phase of the CME LE speed begins, whereas in the right panel, it spans the height range corresponding to the peak CME acceleration.}
\label{fig:all_speed}
\end{figure*}

The left panel of Figure~\ref{fig:all_speed} depicts the LE speed of the selected CMEs with the height of the CME LE. The top and bottom panels of the figure show the LE speed of 7 fast and 7 slow CMEs. In each panel, the legends are ordered in decreasing order of CME speed (from highest to lowest). The error bars are derived by considering a 5\% error in the measurements of the height at each data point (as discussed in Section~\ref{sec:obs3D}) and are represented by transparent shaded regions over the data points, using the same color as the data points. Table~\ref{tab:tab_2} lists the estimated 3D parameters as well as POS projected parameters (from the CDAW catalog on \url{https://cdaw.gsfc.nasa.gov/CME_list/}) for the fast and slow CMEs in the top and bottom panels, respectively.

From the top panel, it is evident that most fast CMEs are first tracked at heights of $\gtrsim3~R_\odot$, except for the events on 2011 March 07, 2012 March 07, and 2012 July 12. In contrast, slow CMEs are typically first detected at a distance of approximately 2 $R_\odot$. The LE speeds at tracked initial heights exceed 2000 $km~s^{-1}$ for three fast CMEs, are about 680 $km~s^{-1}$ for one, and above 1000 $km~s^{-1}$ for the remaining three. In most cases, the LE speed decreases with height, indicating deceleration of the fast CMEs in the coronagraphic field of view, except for the 2012 March 07 and 2012 July 12 events. From the bottom panel of the table, most slow CMEs exhibit initial LE speeds between 200--450 $km~s^{-1}$. An exception is the 2010 June 16 event, which is unusually slow, with an initial speed of about 75 km s$^{-1}$, comparable to the speed at the initial height reported by \citet{Lorenzo2024}. Most selected CMEs exhibit the expected behavior, namely the deceleration of fast CMEs and the acceleration of slow CMEs at larger coronal heights \citep{Vrsanak2007,Sachdeva2015}.

For slow CMEs, the LE speed profile exhibits a two-phase evolution--the speed increases during initial heights, which then approaches an approximately constant value. The beginning of the constant phase spans $\sim$3-4 $R_\odot$ as represented by the grey-shaded region in the figure. In contrast to slow CMEs, fast CMEs (except for the 2012 March 7 and 2012 July 12 CMEs) do not exhibit a consistent speed profile and generally show a decreasing trend in our analysis. For these two exceptionally fast CMEs, the onset of the constant-speed phase occurs over a height range of approximately 4.0-4.5~$R_\odot$, as shown in the grey-shaded region in Figure~\ref{fig:all_speed}.

Further, the LE acceleration of all selected CMEs is depicted in the right column of Figure~\ref{fig:all_speed}. All slow CMEs exhibit a clear acceleration peak (10-145 $m~s^{-2}$ with an average of 65 $m~s^{-2}$) between 2.5--3.5~$R_\odot$, followed by a gradual decline towards a nearly constant value. In contrast, most fast CMEs exhibit a deceleration profile; however, the deceleration is decreasing in magnitude. The two exceptional fast events on 2012 March 7 (peak value of 600 $m~s^{-2}$) and 2012 July 12 (peak value of 250 $m~s^{-2}$) display profiles (evolution and peak) similar to those of slow CMEs. The 2011 March 7 CME shows a deceleration profile with an increase in deceleration. The peak LE acceleration of most fast CMEs likely occurs below 2.5~$R_\odot$, a height range not covered by our tracking, underscoring the importance of observations at lower coronal heights (e.g., \citealt{Zhang2006,Temmer2010}).

The third column of the table lists the average LE speeds of the selected CMEs, derived from linear fits to the de-projected height-time measurements, while the fourth column shows the POS projected speeds from the CDAW catalog. For fast CMEs, we notice the projected speeds are higher or comparable to the de-projected values for CMEs of 2011 March 07, 2011 September 24, 2012 March 07, and 2012 September 28, which contrasts with the general expectation that projected speeds of halo CMEs are underestimated \citep{Sheeley1999,Davies2009,Mishra2013,Mishra2014a,Agarwal2024}. This discrepancy may result from tracking diffuse CME structures, such as shocks or sheaths, in coronagraphic images at larger heights, leading to overestimated projected positions. In contrast to the fast CMEs, the POS projected speed for slow CMEs is less than the de-projected speed, consistent with the expected underestimation, except for one CME of 2010 June 16.

\begin{table*}
    \centering
    \small
    \begin{tabular}{cccccccc}
    \hline
    & Tracked & 3D Speed & Speed in & 3D half & 3D half & Half width in & \\
    Date& height & (Range) [Average] & SOHO & face-on width & edge-on width & (STA) [STB]  & $f(w)$\\
    & range ($R_\odot$) & ($km~s^{-1}$) &  ($km~s^{-1}$)& [$\alpha+\delta$] $(^\circ)$& $[\delta] ~(^\circ)$ & $(^\circ)$ &  \\
    \hline
    \multicolumn{8}{c}{Fast CMEs} \\
    \hline
    2012 Mar 07 & 1.6 - 10 & (2095 - 2750) [2555] & 2684 & 41.5 - 60 & 11.5 - 30 & (103) [63] & 2.13 - 1.58  \\
    2011 Mar 07 & 1.9 - 20.7 & (2254 - 1704) [2029] & 2125 & 44.5 - 56.1 & 14.5 - 26.1 & (66) [50] & 2.02 - 1.67 \\
    2014 Jan 07 & 4.2 - 18 & (2267 - 2002) [2086] & 1830 & 39.3 - 60.5 & 19.3 - 35.5 & (118) [87] & 2.22 - 1.57  \\
    2011 Aug 04 & 4.7 - 17.8 & (1875 - 1307) [1605] & 1315 & 50.6 - 59.5 & 8.6 - 17.5 & (74) [78] & 1.82 - 1.59  \\
    2011 Sep 24 & 3.3 - 17.7 & (1539 - 1435) [1485] & 1915 & 58.7 - 68.3 & 28.7 - 38.3 & (62) [48] & 1.61 - 1.39  \\
    2012 Sep 28 & 3.7 - 17.3 & (1196 - 732) [915] & 947 & 68.7 - 73.4 & 28.7 - 33.4 & (140) [-] & 1.39 - 1.3 \\
    2012 Jul 12 & 1.8 - 15 & (682 - 1128) [1134] & 885 & 46.1 - 62.6 & 21.1 - 37.6 & (49) [-] & 1.96 - 1.52 \\
    \hline
    \multicolumn{8}{c}{Slow CMEs} \\
    \hline
    2010 Apr 03 & 2.1 - 13.7 & (461 - 858) [767] & 668 & 40.1 - 47.2 & 15.1 - 21.7 & (34) [28] & 2.19 - 1.93 \\
    2011 Feb 01 & 1.7 - 13.3 & (408 - 701) [737] & 437 & 29.2 - 46.7 & 9.2 - 26.7 & (60) [57] & 2.78 - 1.94 \\
    2010 Feb 07 & 2.4 - 23.2 & (472 - 405) [489] & 421 & 21.5 - 46.8 & 11.5 - 34.8 & (46) [38] & 3.53 - 1.94 \\
    2010 Apr 08 & 2.1 - 13.8 & (289 - 437) [507] & 264 & 55 - 66.3 & 8 - 16.3 & (45) [42] & 1.7 - 1.45 \\
    2010 May 24 & 1.8 - 14.1 & (273 - 587) [520] & 427 & 35.4 - 61.9 & 10.4 - 36.9 & (51) [45] & 2.41 - 1.53 \\
    2009 Dec 16 & 2.3 - 14.3 & (201 - 456) [397] & 276 & 16.9 - 20.8 & 6.9 - 9.8 & (29) [26] & 4.29 - 3.63  \\
    2010 Jun 16 & 2.4 - 17.5 & (75 - 304) [217] & 236 & 15.7 - 32.1 & 5.7 - 12.1 & (28) [31] & 4.55 - 2.59 \\
    
    \hline 
    \end{tabular}
    \caption{The table lists the tracked height range, speed at the initial and last tracked heights in (Range) along with de-projected speed from the linear fitting of height-time measurements in [Average], POS projected constant speed from SOHO coronagraphic observations, as reported in the CDAW catalog, half face-on width (at the initial and last tracked heights), half edge-on width (at the initial and last tracked heights), projected half angular width from \textit{STEREO-A} and \textit{STEREO-B} in (STA) and [STB], as reported in SEEDS catalog, and the ratio of leading edge speed to the full cone model (at the initial and last tracked heights) expansion speed for fast and slow CMEs in the top and bottom panels, respectively.}
    \label{tab:tab_2}
\end{table*}

\subsection{Angular width of CMEs}{\label{sec:angular_width}}

The face-on and edge-on angular widths of a CME are linked to CME spatial extent along different directions. We estimate the half face-on angular width ($\alpha+\delta$) along the toroidal direction (along the axis of torus) and half edge-on angular width ($\delta$) along the poloidal direction (in planes orthogonal to the toroidal axis) of CMEs by utilizing their half angle ($\alpha$) between the conical legs and the aspect ratio ($\kappa=sin(\delta)$) of the hollow croissant CME \citep{Thernisien2011}. For all selected CMEs, except for the 2010 June 16 event, the half angle obtained from GCS fitting remains approximately constant from the initial to the final tracked heights. This implies that the half angle has fully evolved at the lowest coronal heights considered, where our events lack coronagraphic observations. For the CME of 2010 June 16, a change in $\alpha$ by about $10^\circ$ is noticed within 5 $R_\odot$, and then it remains constant to the last tracked height.

\begin{figure}
    \centering
    \includegraphics[scale= 0.75,trim={0cm 0cm 0cm 0cm},clip]{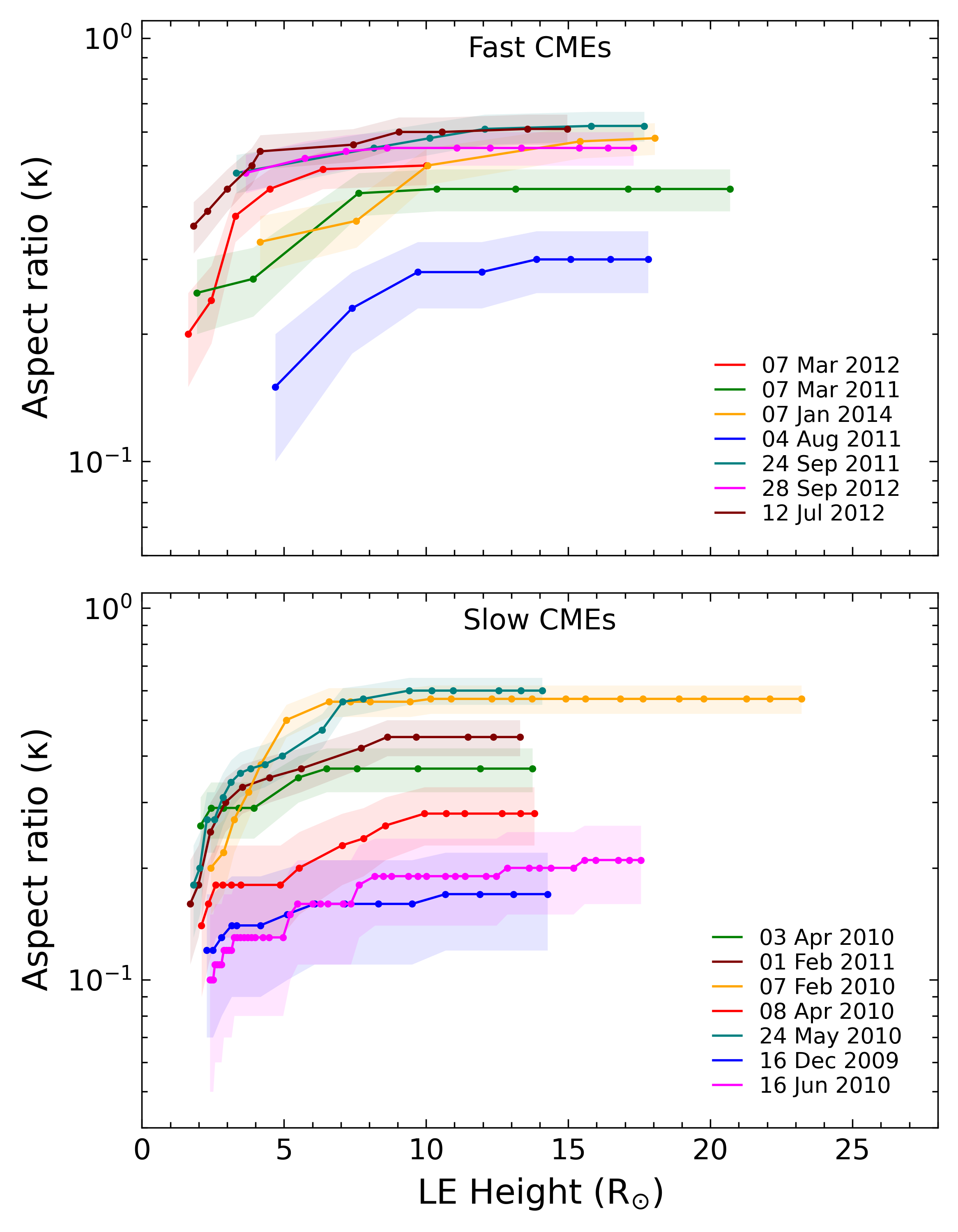}
    \caption{The top and bottom panels show the evolution of the aspect ratio of the fast and slow CMEs with the height of the CME LE. The legends indicate the CME dates, ordered in decreasing order of CME speed (from highest to lowest). The error bars are represented by the transparent shaded regions over the data points, in the same color as the data points.}
    \label{fig:kappa_evol}
\end{figure}

The aspect ratio of a hollow‐croissant CME can govern its 3D angular widths, radius, and expansion speeds; therefore, examining its evolution is particularly insightful. Figure~\ref{fig:kappa_evol} depicts the evolution of the aspect ratio with the height of the CME LE for fast CMEs in the top panel and for slow CMEs in the bottom panel. The legends are arranged in decreasing order of CME speed, from highest to lowest speed. The figure indicates that $\kappa$ initially increases and subsequently reaches a nearly constant value, consistent with the indicated trend in previous studies \citep{Cremades2020,Agarwal2024}. For fast CMEs, $\kappa$ generally increases rapidly at lower heights, and most of them exhibit reaching a constant aspect ratio (i.e., saturation) beyond 5-7 $R_\odot$, with $\kappa$ approaching nearly constant values in the range of 0.4-0.6, suggesting a transition toward self-similar expansion at larger heights.

For slow CMEs, the aspect ratio also increases with height but remains systematically smaller than that of fast CMEs. The growth of $\kappa$ is more gradual, and the saturation values are typically lower, lying in the range of 0.15-0.6, and takes place beyond 8-10 $R_\odot$. The study of \citet{Agarwal2024,Mishra2026} proposed a three-phase evolution in which $\kappa$ increases within a few solar radii, then remains constant over certain heights, followed by a decrease in the IP medium. We note that on average, the aspect ratio increases by about 73\% for fast CMEs and 128\% for slow CMEs over their tracked height ranges ($\sim$ 2-20 $R_\odot$) in the middle and outer corona. The larger relative increase in aspect ratio observed for slow CMEs than for fast CMEs could be explained if the fast CMEs underwent a much stronger expansion at lower coronal heights, which were missed in our coronagraphic observations. Several of the CMEs analyzed in this study (2009 December 16, 2010 April 03, 2010 April 08, 2010 May 24, 2010 June 16, 2011 August 04, 2012 July 12, and 2012 September 28) were also examined by \citet{Temmer2021}. In their study, the CMEs on 2009 December 16, 2010 May 24, 2011 August 04, and 2012 September 28 have noticeable differences in the aspect ratio compared to our estimates. While it appears that such differences in the magnitude of the aspect ratio may not affect the overall evolutionary trend of $\kappa$, the qualification of exact errors arising from various sources in each event is difficult to quantify accurately \citep{Thernisien2009,Pluta2019,Cremades2020}. This will lead to deviations in the inferred flux-rope radius and in the face-on and edge-on angular widths. Since our primary focus is on the evolutionary trends in angular width and associated parameters, including spatial dimensions and expansion speeds, we do not expect uncertainties in the aspect ratio at a given height to significantly affect the derived trends.

It is interesting to note that although the angular width (2$\delta$) of the individual conical legs in GCS assumed flux rope shows variations with height, the angular separation (2$\alpha$) between two conical legs remains almost constant for the majority of CMEs. This implies that variations in the face-on angular width with height are governed by changes in the edge-on angular width, through the change of $\delta$.

The fifth and sixth columns of Table~\ref{tab:tab_2} list the estimated half face-on and edge-on angular widths of the selected CMEs at their initial and final tracked heights. For fast CMEs, the half face-on and half edge-on width increase on average by 29\% and 79\%, respectively, from the initial to the last tracked height. Similarly, for slow CMEs, the corresponding increases are 60\% and 135\%, indicating a stronger expansion in comparison to fast CMEs. This suggests that fast CMEs may already be overexpanded at initial heights \citep{Patsourakos2010a,Morosan2022}. For the CMEs on 2012 Jul 12, 2010 Jun 16, 2010 Apr 03, 2012 Sep 28, 2010 May 24, and 2009 Dec 16, the half face-on and half edge-on angular widths estimated in this study agree with those reported by \citet{Temmer2021} within $10^\circ$. While for the 2010 Apr 08 and 2011 Aug 04 CMEs, the corresponding differences are approximately $20^\circ$ in the half face-on angular width and $10^\circ$ in the half edge-on angular width, respectively. Such differences are likely due to user-dependent variability in the GCS fitting procedure.

We compare the face-on angular widths (fifth column) of CMEs with the POS projected angular width (seventh column, Table~\ref{tab:tab_2}) in \textit{STEREO-A} and \textit{STEREO-B} observations, taken from the SEEDS catalog (\url{http://spaceweather.gmu.edu/seeds/secchi.php}). A dashed line indicates CMEs not detected by the SEEDS automatic algorithm. For most CMEs, the projected half angular widths from \textit{STEREO-A} and \textit{STEREO-B} are similar, except for the 2012 March 07 and 2014 January 07 events, where the SEEDS algorithm likely overestimated widths in \textit{STEREO-A} by including non-CME structures. For the CMEs on 2012 March 07, 2014 January 07, 2011 September 24, 2012 September 28, 2010 April 03, and 2010 April 08, the half face-on angular width at the final tracked height differs by more than $15^\circ$ from the corresponding projected angular width observed from either STEREO-A or STEREO-B. Such differences likely arise from CMEs' tilt relative to the ecliptic plane.

\subsection{Radial and lateral expansion of CMEs}{\label{sec:radlatexp}}

 Thus, utilizing the GCS model-derived evolution of radial and lateral dimension, in this section, we estimate the radial expansion speed (${V_{exp\_rad}}$) along the CME propagation direction and lateral expansion speed (${V_{exp\_lat}}$) along the perpendicular to CME propagation direction in the plane of CME propagation. In the GCS model framework, of the two lateral expansion speeds, the one in the plane perpendicular to CME propagation equals the radial expansion speed. We focus on comparing radial and lateral expansion speeds, which are in the same plane as the CME propagation. Whenever we refer to lateral expansion speed, we mean the one in the plane of CME propagation.

As described earlier, the face-on angular width along the toroidal direction serves as a proxy for the lateral dimension in the plane of CME propagation. Likewise, the edge-on angular width along the poloidal direction represents the lateral dimension perpendicular to the plane of CME propagation. Owing to the GCS geometry, in which the toroidal cross-sections are perfect circles, the lateral dimension perpendicular to the plane of CME propagation is identical to the radial dimension along the CME propagation direction.

\begin{figure*}
\centering
\includegraphics[scale= 0.06,trim={0cm 0cm 0cm 0cm},clip]{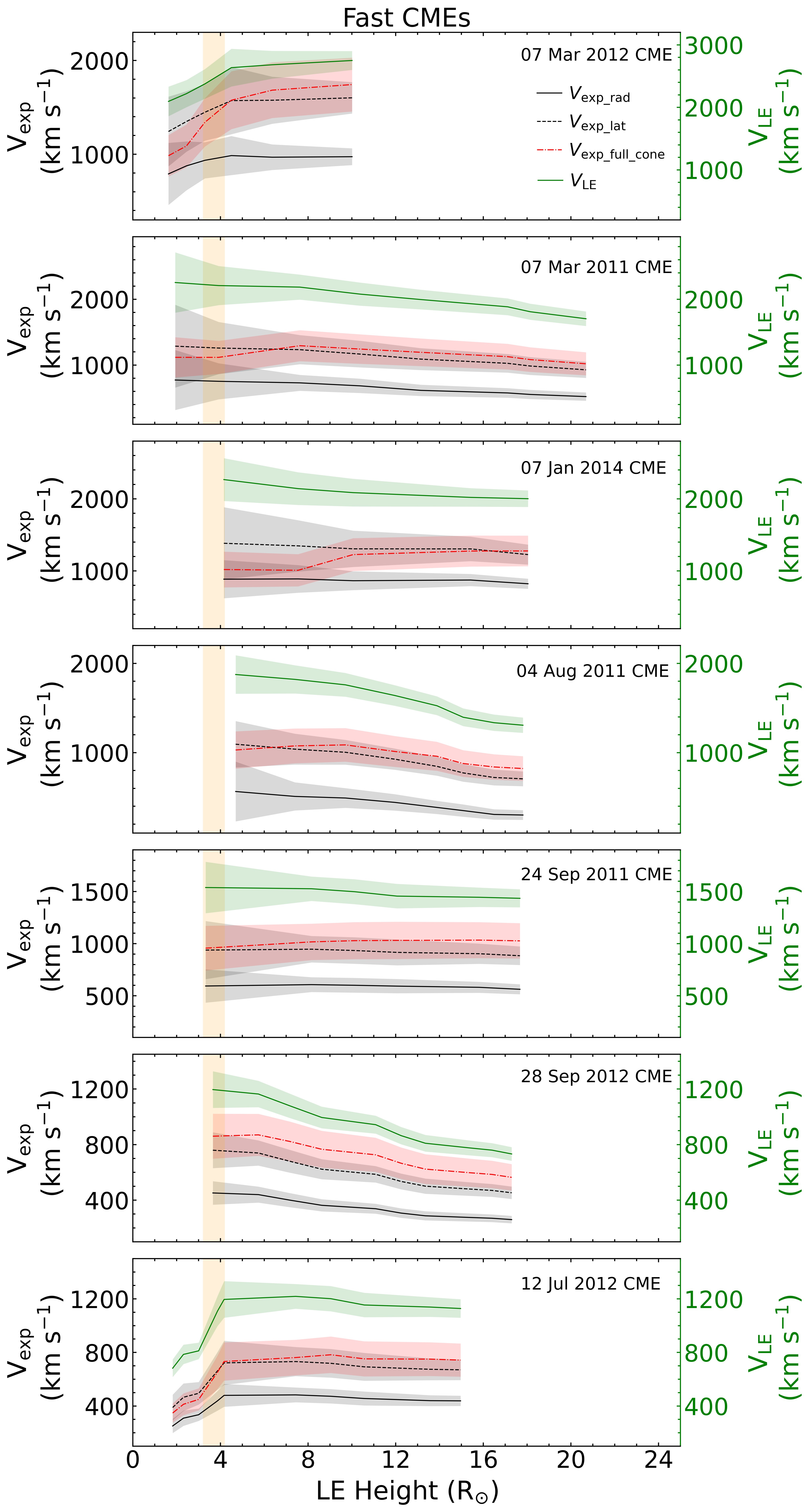}
\includegraphics[scale= 0.06,trim={0cm 0cm 0cm 0cm},clip]{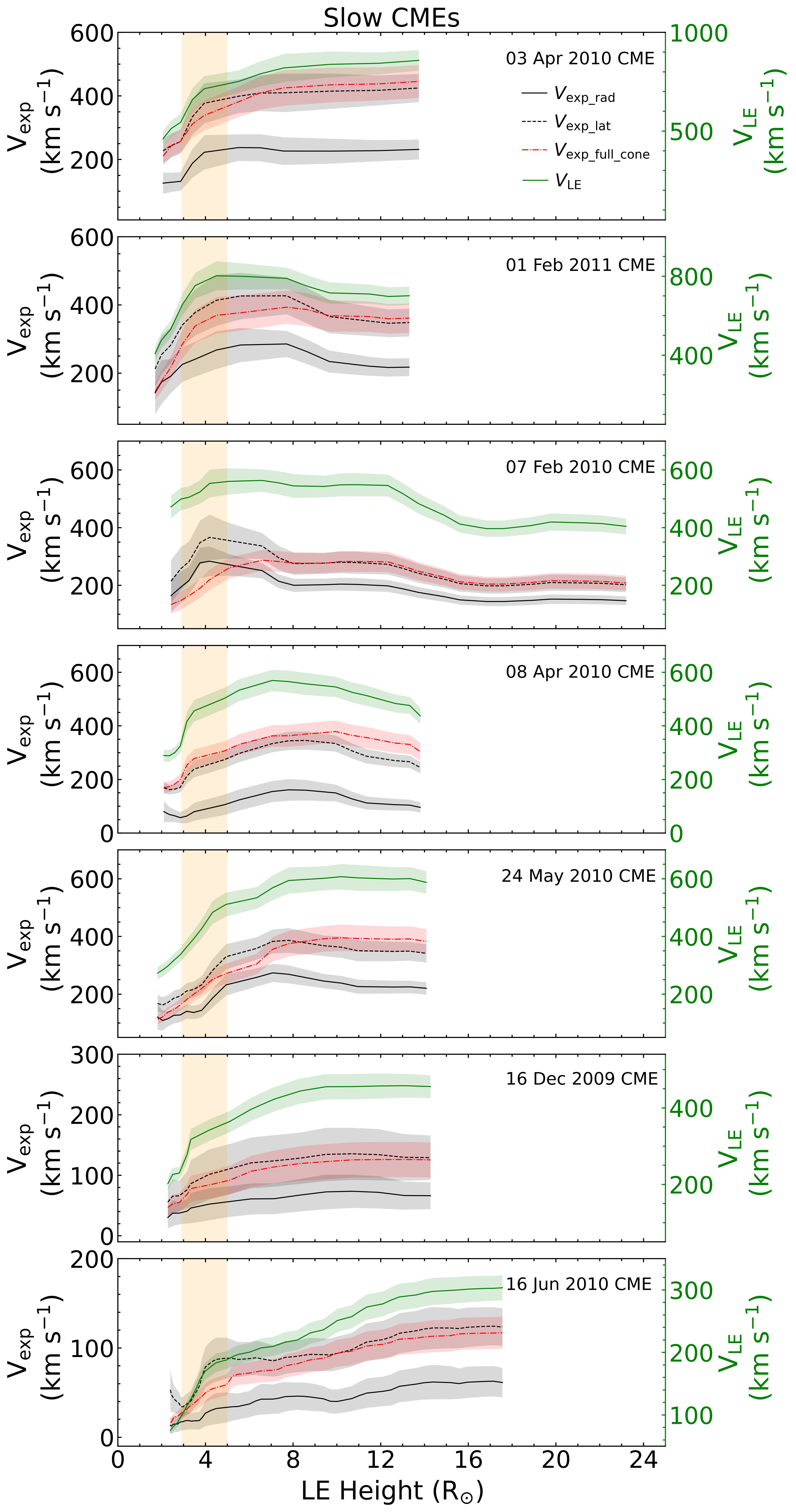}
\caption{The left and right panels depict the expansion speeds of fast and slow CMEs with the height of the CME LE. The radial expansion speed, lateral expansion speed, and the expansion speed from the full cone model are plotted on the left y-axis and are represented by black solid, black dashed, and red dash-dotted lines, respectively. The CME LE speed is plotted on the right y-axis as a green solid line. The CMEs are ordered in decreasing CME LE speed (from highest to lowest). Transparent shaded regions over the curves represent the error bars and are shown in the same color as the corresponding curves. The orange-shaded region marks the approximate height range where the constant phase of the expansion speed begins.}
\label{fig:expansion_speed}
\end{figure*}

Our analysis focuses on comparing the radial and lateral expansion speeds that lie in the same plane as the CME propagation. Unless stated otherwise, references to estimated lateral expansion speed in the present study hereafter correspond to the component within the plane of CME propagation. Using the GCS model–derived evolution of the radial and lateral dimensions, we estimate in this section the radial expansion speed (${V_{exp\_rad}}$) along the CME propagation direction and the lateral expansion speed (${V_{exp\_lat}}$) within the plane of CME propagation. The radius ($R_{rad}$) of the GCS-modeled CME flux rope is estimated by the $\kappa$ and height ($h_f$) of the CME LE (front), as listed in Equation (i) of Table~\ref{tab:gcs_parameter}. We note that one can also write this as $R_{rad} = {\kappa}~h_c$, where $h_c$ ($OC_1$) is the height of the CME center. The estimated radial dimension (diameter of the flux rope), at a height of around 10 $R_\odot$, ranges from 4–8 $R_\odot$ for fast CMEs and 3–8 $R_\odot$ for slow CMEs (e.g., \citealt{Cremades2020,Zhuang2023,Agarwal2024}).

Previous studies employing the GCS model have primarily focused on estimating the CME center speed and radial expansion speed \citep{Scolini2019,Zhuang2023,Agarwal2024}. In this study, we additionally derive the lateral expansion speed using half of the CME lateral extent ($R_{\rm lat}$) obtained from the GCS model, as described in Appendix~\ref{sec:appendix}. In an earlier study, \citet{Cremades2020} also estimated the lateral expansion speed of CMEs using the temporal evolution of GCS-derived parameters. Using our method, we find that the lateral dimension ($2R_{\rm lat}$), measured in the plane of the CME propagation direction at a height of approximately 10~$R_\odot$, ranges from 10--14~$R_\odot$ for fast CMEs and 8--12~$R_\odot$ for slow CMEs.

The radial and lateral expansion speeds are estimated using the moving box linear fit technique to the radius ($R_{rad}$) and half of its lateral extent ($R_{lat}$) of the CME flux rope \citep{Agarwal2024}. Figure~\ref{fig:expansion_speed} shows the radial (solid black) and lateral (dashed black) expansion speeds on the left y-axis as a function of CME LE height, with the left and right panels representing fast and slow CMEs, respectively. The LE speed (solid green) is plotted on the right y-axis. Error bars are shown as transparent shaded regions in the same color as the corresponding curves and are derived through error propagation, using the uncertainties assigned to the GCS parameters as described in Section~\ref{sec:obs3D}. We note that the radial expansion speed is lower and appears to be a scaled-down version of the lateral expansion speed, for both fast and slow CMEs, indicating asymmetric expansion during their propagation in the IP medium \citep{Patsourakos2010,Veronig2018,Cremades2020}.

For slow CMEs, the radial and lateral expansion speeds exhibit similar profiles, displaying a two-phase evolution within our observed heights: the speed increases during initial heights, which then approaches an approximately constant value. Although the onset of the constant phase varies from event to event, it occurs within the height range of $\sim$3--5 $R_\odot$. This height range is highlighted by the orange shaded region in Figure~\ref{fig:expansion_speed}. In contrast to slow CMEs, fast CMEs do not exhibit a consistent trend, except for the two events of 2012 March 07 and 2012 July 12, which show profiles similar to those of slow CMEs. For these two events, the onset of the approximately constant expansion phase occurs within a height range of $\sim$3--4 $R_\odot$, which is highlighted by the orange-shaded region. The other fast CMEs display either decreasing or constant expansion profiles. It is possible that the peak phase is often unobserved for the selected fast CMEs, emphasizing the importance of tracking them from lower coronal heights \citep{Temmer2010}. The study of \citet{Khuntia2024} reported thermal state transition (heat release to heat absorption) at heights around 3--7 $R_\odot$, which overlaps with the height range where we observe the transition to a constant expansion phase, suggesting a possible physical connection between CME thermodynamics and expansion behavior.

\subsection{GCS-Derived Radial and Lateral CME Expansion versus Full Cone Model Expansion}{\label{sec:fullexp}}

We estimate CME expansion using the full cone model \citep{Gopalswamy2009,Michalek2009,Gopalswamy2012,Makela2016} and compare it with the radial and lateral expansion speeds derived from the GCS model. The full cone model assumes equal lateral dimensions both in the plane of CME propagation and in the plane perpendicular to the CME propagation direction. It further assumes that the radial dimension ($2R_{rad}$) and lateral dimension ($2R_{lat}$) within the plane of CME propagation are identical, implying a single expansion speed applicable along all three mutually perpendicular directions. This expansion speed is estimated using the half-base length of the cone as follows:

\begin{equation}
V_{LE} = f(\omega)V_{exp\_full\_cone} = (1 + cot\omega)V_{exp\_full\_cone}
\end{equation}

\begin{equation}\label{equ:f(w)exp_le}
V_{exp\_full\_cone} = \frac{1}{f(\omega)}V_{LE}
\end{equation}

Earlier studies assumed a constant $\omega$, corresponding to the projected half angular width of the cone in the full cone model \citep{Gopalswamy2009,Michalek2009,Gopalswamy2012,Makela2016}. In the present study, we instead use the de-projected, time-varying half face-on angular width of the hollow croissant-shaped CME as $\omega$ to estimate the de-projected expansion speed from the full cone model, as given in Equation~\ref{equ:f(w)exp_le}, where $f(\omega)$ is $(1 + cot\omega)$. We use the face-on angular width because it represents the full angular extent between the two outer flanks of the croissant-shaped shell, which is the closest geometric analog to the opening angle of a 3D cone. The resulting full cone model expansion speed ($V_{{exp\_full\_ cone}}$) is shown in Figure~\ref{fig:expansion_speed} by the red dash-dotted line, with error bars represented by transparent red shaded regions around each data point.

From the left panel (fast CMEs), the full cone model expansion speed is generally comparable to the lateral expansion speed at all heights, except for the 2012 March 07 and 2014 January 07 events. For these two CMEs, the full cone expansion speed matches the radial expansion speed at lower heights and transitions to the lateral expansion speed at higher heights. From the right panel (slow CMEs), the full cone expansion speed for the 2010 April 03, 2010 April 08, and 2009 December 16 events closely follows the lateral expansion speed at all heights. For the remaining slow CMEs, it agrees with the radial expansion speed at lower heights and with the lateral expansion speed at higher heights. Overall, these results indicate that the expansion speed derived from the full cone model primarily represents the lateral expansion of CMEs, especially at larger heights. The limitations of the full cone model are discussed in Section~\ref{sec:resdis}.

The last column of Table~\ref{tab:tab_2} lists the values of $f(\omega)$ at the initial and final tracked heights, with the top and bottom panels corresponding to fast and slow CMEs, respectively. On average, $f(\omega)$ decreases by about 18\% for fast CMEs and 28\% for slow CMEs from the initial to the final tracked height. This behavior indicates that empirical relations assuming a fixed $f(\omega)$ at different heights of an evolving CME \citep{Gopalswamy2009} have inherent limitations. Moreover, adopting a constant $f(\omega)$ for all CMEs, as assumed in some earlier studies \citep{Dallago2003,Schwenn2005}, may not be justified. Because CME expansion evolves with propagation. To further explore CME dynamics, the following section examines the correlations between different CME speeds (expansion and LE) and their angular widths.

\section{Relation between speeds and angular widths of CMEs}{\label{sec:linrel}}

\begin{figure*}
\centering
\includegraphics[scale=1.8,trim={0.26cm 0.26cm 0.25cm 0.25cm},clip]{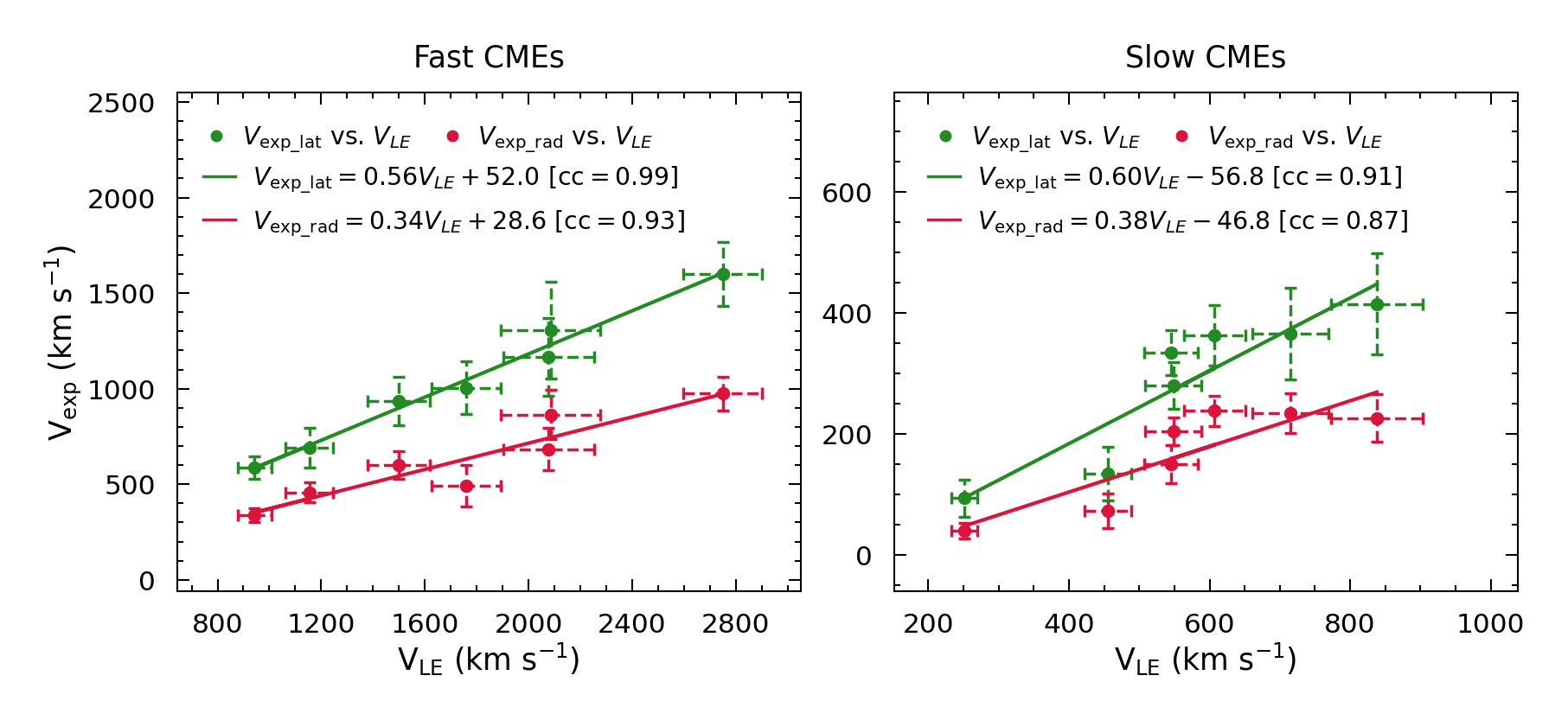}\\
\includegraphics[scale=1.8,trim={0.25cm 0.25cm 0.24cm 0.41cm},clip]{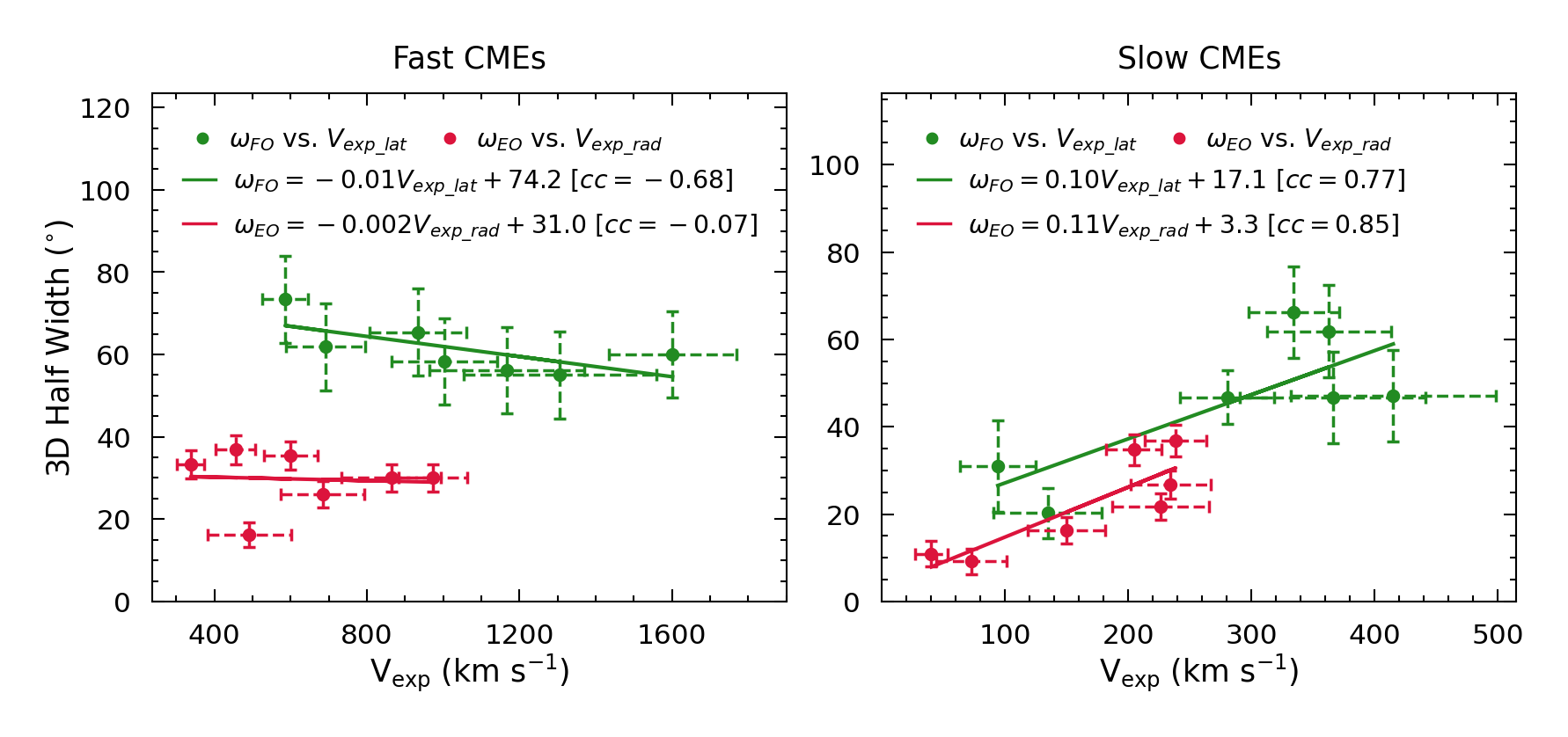}\\
\includegraphics[scale= 1.8,trim={0.25cm 0.25cm 0.24cm 0.41cm},clip]{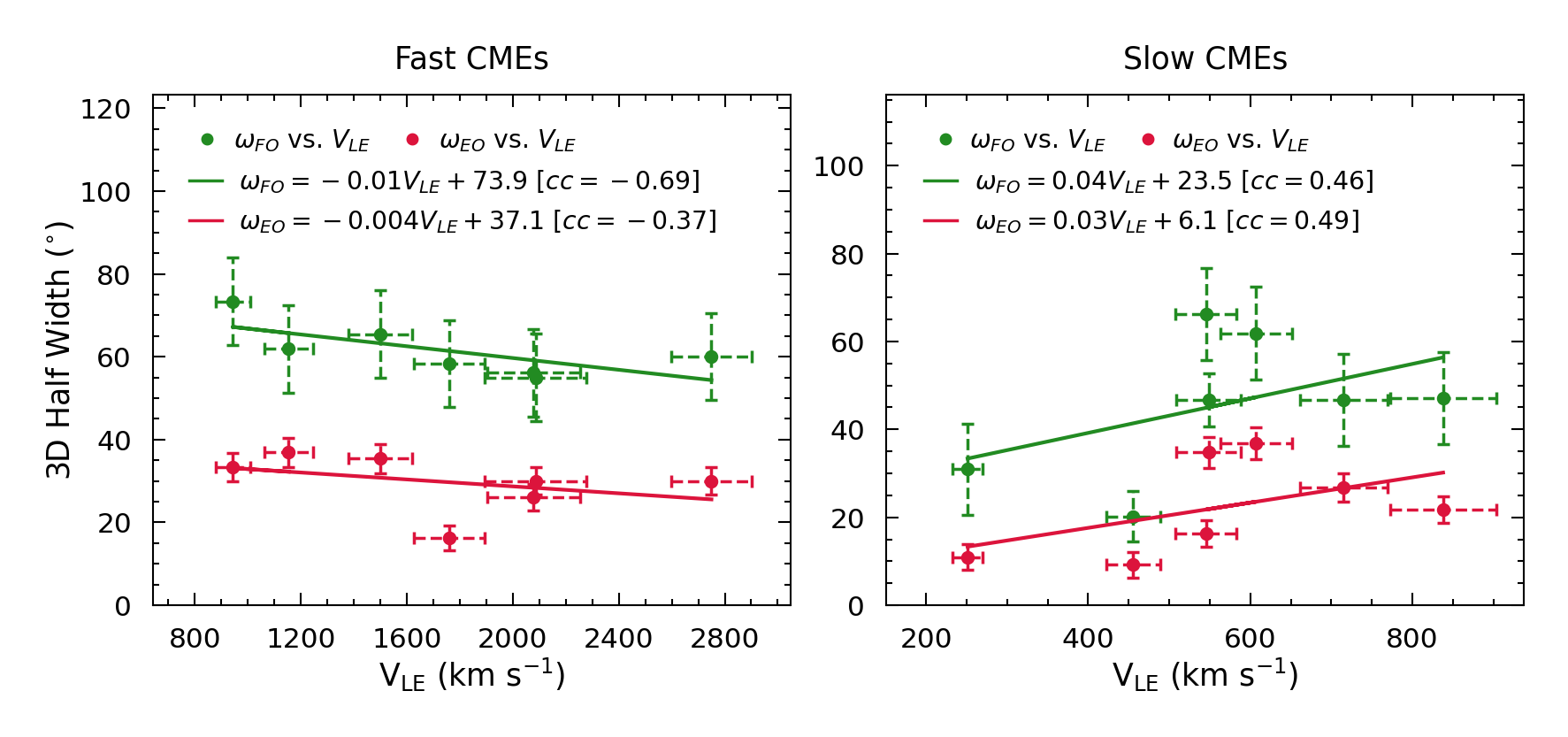}
\caption{The top panel shows the correlation of both expansion speeds (radial and lateral) with the LE speed. The middle panel shows the correlation between the half face-on angular width and the lateral expansion speed, as well as between the half edge-on angular width and the radial expansion speed. The bottom panel shows the correlation between the angular widths (edge-on and face-on) and the LE speed. The left and right columns correspond to fast and slow CMEs, respectively. Each data point represents the plotted parameter for a CME, computed at the height of 10~$R_\odot$. The dashed line represents the error bars.}
\label{fig:width_speed}
\end{figure*}

We examine whether CMEs with faster radial and lateral expansion also propagate faster by correlating their expansion speeds with the LE speed. In the top panel of Figure~\ref{fig:width_speed}, the correlation between lateral expansion speed and LE speed is shown in green, while that between radial expansion speed and LE speed is shown in red, for fast CMEs (left column) and slow CMEs (right column). For each CME, a single representative data point is used for each parameter, taken at the height around $10~R_\odot$. CME parameters at $10~R_\odot$ enables a consistent comparison between fast and slow CMEs, since key properties, such as aspect ratio, expansion speeds, LE speed, and 3D angular width, are reasonably evolved by this height and remain approximately constant beyond this distance (Section~\ref{sec:obs3D}).

In our analysis, correlation coefficients ($cc$) in the ranges $0.6 < cc\leq 1$, $0.4 < cc \leq 0.6$, and $0 < cc \leq 0.4$ are classified as strongly, moderately, and weakly correlated, respectively. Such a range of $cc$ values and the classification of correlation strength are also considered in \citet{Mishra2024}. For both fast and slow CMEs, the radial and lateral expansion speeds exhibit a strong positive correlation with the LE speed, as indicated by the correlation coefficients and linear fit equations shown in the figure. This implies that CMEs with higher LE speeds tend to exhibit greater expansion rates in both radial and lateral directions at 10 $R_\odot$, consistent with the findings of \citet{Gopalswamy2009}, which indicates enhanced internal magnetic pressure within the flux rope and/or between its conical legs relative to the surrounding medium.

We also examine whether CMEs that expand more rapidly are indeed wider. To investigate this, we derived the correlation between the half face-on angular width ($\omega_{FO}$) and the lateral expansion speed (shown in green), and between the half edge-on angular width ($\omega_{EO}$) and the radial expansion speed (shown in red), as presented in the middle panel of the figure. For fast CMEs, the half face-on angular width along a toroidal direction shows a strong negative correlation with the lateral expansion speed. This indicates that fast CMEs with higher lateral expansion speeds are not necessarily wider in the lateral direction at a height of around $10~R_\odot$. This behavior likely reflects that some fast CMEs with large lateral expansion speeds have not yet completed their expansion phase; once the expansion saturates, such CMEs are expected to exhibit larger angular widths and lower expansion speeds. In contrast, the correlation between the half edge-on angular width along the poloidal direction and the radial expansion speed is weak. Its weakness likely arises because the aspect ratio ($\kappa$) varies only marginally among most fast CMEs at heights $10~R_\odot$. Unlike the fast CMEs, slow CMEs show a strong positive correlation between $\omega_{FO}$ and $V_{exp\_lat}$, suggesting that their expansion is nearly exhausted at heights around $10~R_\odot$. As expected, slow CMEs also exhibit a strong positive correlation between $\omega_{EO}$ and $V_{exp\_rad}$.

We also investigate whether wider CMEs exhibit higher LE speeds. To investigate this, we derived the correlation of both angular widths with the LE speed, as shown in the bottom panel of the figure. For fast CMEs, the $\omega_{FO}$ exhibits a strong and negative correlation with the LE speed, indicating that narrower CMEs among the selected fast CMEs have a higher LE speed at height around 10 $R_\odot$. This could be due to wider fast CMEs experiencing a larger drag force \citep{Cargill2004}. The $\omega_{EO}$ exhibits a negative but weak correlation with the LE speed. For slow CMEs, both angular widths show moderate but positive correlations with the LE speed. In contrast to these results, earlier studies by \citet{Gopalswamy2009} and \citet{Gopalswamy2014} reported a positive correlation between CME width and LE speed, likely because fast and slow CMEs were treated as a single population \citep{Vrsanak2004}. Overall, we find that fast and slow CMEs follow fundamentally different evolutions in terms of expansion versus propagation. However, because of the limited sample of CMEs in each CME category, these results may be affected by sample bias. A larger CME sample in future studies will be important to further validate and generalize these findings.

\section{Results and Discussion}{\label{sec:resdis}}

\subsection{Expansion characteristics and asymmetry}

The present study focuses on the asymmetric expansion of CMEs in the radial and lateral directions up to coronagraphic heights, along with the evolution of their face-on angular width along the toroidal direction and edge-on angular width along the poloidal direction. To better understand the evolution of these characteristics of CMEs, we analyzed 14 CMEs and categorized them as 7 fast or 7 slow based on their LE speed (as described in Section~\ref{sec:obs3D}). We find that, regardless of their speed category (fast or slow), the lateral expansion speed exceeds the radial expansion speed. This asymmetric expansion is consistent with earlier observational and modeling studies, reflecting the combined influence of internal magnetic pressure and interaction with the surrounding corona and heliosphere \citep{Patsourakos2010,Cabello2016,Cremades2020}.

Slow CMEs show a coherent two-phase expansion behavior, with both radial and lateral expansion speeds increasing at lower heights and beginning to approach a constant value around $\sim3$–5~$R_\odot$. In contrast, fast CMEs exhibit more diverse expansion profiles: while a few events show a similar transition to a constant expansion phase, most display either decreasing or nearly constant expansion speeds within the observed height range. This suggests that the main expansion phase of fast CMEs might occur at heights below those sampled here, suggesting their rapid early evolution. However, for both fast and slow CMEs, the expansion speeds become nearly constant at higher heights, consistent with the findings of \citet{Balmaceda2020}.

In our study, the average ratio of lateral expansion speed to radial expansion speed at the last tracked height is 1.7 for fast CMEs and 1.8 for slow CMEs, both of which are nearly equal to the ratio of 1.6 reported in the study by \citet{Cremades2020}. The average ratio of LE speed to lateral expansion speed at the last tracked height is 1.71 for fast CMEs, closely matching the value of 1.76 reported by \citet{Dallago2003}, and 2.22 for slow CMEs, which is comparable to the value of 2.34 reported by \citet{Michalek2009}.

The observed evolution of angular widths reinforces the contrasting expansion histories of fast and slow CMEs. The percentage increase in the half face-on (along the toroidal direction) and half edge-on (along the poloidal direction) angular widths from the initial to the final tracked heights is, on average, about 2 times and 1.7 times larger, respectively, for slow CMEs than for fast CMEs. This stronger angular growth in slow CMEs is consistent with their sustained expansion speed. The average of the half face-on to edge-on angular widths ratio at the last tracked height is 2.1 for fast CMEs and 2.3 for slow CMEs, both of which are higher than the value of 1.8 reported by \citet{Cremades2020}.

The evolution of the aspect ratio ($\kappa$) provides further insight into expansion behavior. We found that, for both fast and slow CMEs, $\kappa$ increases at lower heights followed by a tendency toward saturation at larger heights \citep{Cremades2020}. Fast CMEs exhibit a sharper increase below 5-7 $R_\odot$ and attain higher, nearly constant (0.4-0.6) $\kappa$ values. In contrast, slow CMEs show a more gradual increase up to 8–10 $R_\odot$ and reach systematically lower quasi-constant (0.15-0.6) values. On average, slow CMEs show a larger relative increase in $\kappa$ from their initial to final tracked heights than fast CMEs. This supports the idea that fast CMEs undergo stronger expansion at lower heights, whereas slow CMEs exhibit a more gradual expansion extending over a larger radial range.

In our GCS fitting of CME events, a smaller change is noted for the half angle ($\alpha$) while a significant change in $\kappa$ for the majority of CMEs as they evolve at higher heights. Although this trend is evident in our analysis, its physical significance remains unclear because GCS fitting involves subjective adjustments to model parameters. Therefore, GCS reconstructions of the same CME events by multiple independent users would be needed to quantify the effect of user-dependent bias. We also acknowledge that the GCS-derived fitting parameters are subject to uncertainties arising from inherent degeneracies in multiple parameters, which are difficult to minimize. In this study, these uncertainties are minimized by carefully selecting CME events that were simultaneously observed by multiple spacecraft with favorable angular separation \citep{Verbeke2023,Lyu2023,Nikou2025}.

\subsection{Full cone model versus GCS-derived expansion}

We also estimate the expansion speed of selected CMEs using the full cone model \citep{Gopalswamy2009,Michalek2009} and compare it with those obtained from the GCS model (Section~\ref{sec:radlatexp}). For most CMEs, regardless of speed category, the expansion speed from the full cone model closely matches the lateral expansion speed at all heights. However, in some cases, it follows the radial expansion speed at lower heights and the lateral expansion speed at higher heights. These results suggest that the full cone model expansion speed does not accurately represent the radial expansion speed, particularly at higher heights, and can alter the findings of previous studies that employed the full cone model \citep{Michalek2009,Makela2016,Balmaceda2020,Mishra2021a}.

We employ the face-on angular width, which best represents the CME’s 3D cone in the plane of propagation, to estimate the expansion speed using the full cone model. In the limiting case of the GCS model, setting the half angle to zero reduces the geometry to an ideal cone, for which the edge-on angular width is more appropriate. Therefore, expansion speeds derived from the full cone model should be interpreted with caution, as their validity depends on the CME’s global structure and its deviation from the ideal cone over different height ranges. Although the tilt angle ($\gamma$) is not used directly in our analysis for estimating angular widths or expansion speeds, uncertainties in GCS model fitting of the $\gamma$ can influence other derived three-dimensional parameters, particularly the half-angle between the conical legs ($\alpha$) and the aspect ratio \citep{Verbeke2023}. These uncertainties primarily affect the absolute values of the parameters, but not their overall evolutionary trends.

Moreover, the average value of ${f(\omega)}$ (the ratio of LE speed to expansion speed from the full cone model) decreases by about 18\% for fast CMEs and 28\% for slow CMEs from the initial to the final tracked height. This larger percentage decrease in slow CMEs can be attributed to their higher rate of increase in angular width compared to fast CMEs, or to the overexpansion of fast CMEs at lower heights. These findings suggest that relying on a single, fixed value of ${f(\omega)}$ is insufficient for accurately estimating CME lateral expansion speed, even though earlier studies have established several empirical relations between LE speed and lateral expansion speed \citep{Dallago2003,Schwenn2005,Michalek2009}.

\subsection{Angular width evolution and projection effects}

Except for one event, all selected CMEs have a sufficiently large angular separation between their propagation direction and the spacecraft position, allowing better estimates of their 3D parameters \citep{Shen2013}. Despite this favorable viewing geometry, we find that, for most fast CMEs, the projected half-angular widths measured from the twin \textit{STEREO} observations (obtained through the automated procedure of the SEEDS catalog) differ from their de-projected half face-on angular widths. This indicates that the projected half-angular widths do not necessarily represent the true de-projected angular widths. Such differences may arise from the tilt of the CME axis relative to the ecliptic plane, as well as limitations of automated width-detection algorithms \citep{Agarwal2024}. The results underscore the importance of 3D reconstruction for accurately characterizing CME angular extent, especially when relating width to kinematic properties.

\subsection{Distinct kinematic evolution of fast and slow CMEs}

Our analysis reveals clear and systematic differences in the kinematic evolution of fast and slow CMEs. We note that most fast CMEs show deceleration and most slow CMEs exhibit gradual acceleration, which is consistent with earlier statistical studies of CME-solar wind interaction \citep{Vrsanak2010,Sachdeva2015}. We also note that the de-projected LE speed of slow CMEs is more than their projected speed and is consistent with earlier findings (e.g., \citealt{Sheeley1999,Davies2009,Mishra2013}). However, for fast CMEs, we observe a deviation from this behavior, likely due to the tracking of other CME substructures, such as shocks or sheaths, in the projected coronagraphic observations.

We note that slow CMEs exhibit a two-phase evolution of the LE speed profile, where the speed initially increases rapidly and then becomes nearly constant. The beginning of the constant phase spans the height range of $\sim 3-4~R_\odot$. The LE acceleration peak ranges from 10-145 $m~s^{-2}$ and occurs at heights of $\sim 2.5-3.5~R_\odot$. In contrast, most fast CMEs exhibit a decreasing speed profile for the LE, with peak acceleration likely occurring at heights below those accessible in our dataset ($<2.5~R_\odot$). This highlights the importance of low-coronal observations for capturing the early dynamics of fast CMEs \citep{Zhang2006,Temmer2010}.

The average peak LE acceleration for slow CMEs is 65 $m~s^{-2}$, whereas previous studies reported LE peak accelerations around 800 $m~s^{-2}$ (e.g., \citealt{Vrsanak2007,Bien2011}). This indicates that the acceleration peak identified here likely corresponds to the residual acceleration occurring at larger heights, rather than the main acceleration peak, which typically forms lower in the corona \citep{Zhang2006}. Consistent with this, \citet{Temmer2010} found that CME peak acceleration generally occurs very close to the source region, at low coronal heights. In our study, the main acceleration peak is likely missed because the initial tracked heights for fast and slow CMEs are around 3~$R_\odot$ and 2~$R_\odot$, respectively. Therefore, low-corona ($\gtrsim 1.05~R_\odot$) observations are crucial for a better understanding of the early evolution and driving mechanisms of CMEs.

\subsection{Coupling between expansion, propagation, and angular width}

Our study shows that, at 10 $R_\odot$, slow CMEs exhibit a strong correlation between radial expansion along the CME propagation direction and the edge-on angular width (along the poloidal direction), as well as between lateral expansion perpendicular to the CME propagation direction and the face-on angular width (along the toroidal direction). This implies that expansion is largely complete by $\sim 10~R_\odot$ and that wider CMEs are indeed those that expanded more. In contrast, for fast CMEs, a negative correlation is observed between the face-on angular width and the lateral expansion speed, suggesting that some fast CMEs with larger expansion speeds have not yet completed their expansion phase, while the half edge-on angular width is poorly correlated with the radial expansion speed.

Furthermore, narrower fast CMEs tend to propagate with higher LE speeds, likely experiencing weaker drag forces compared to wider CMEs, while the half edge-on angular width is poorly correlated with the LE speed. In the case of slow CMEs, both angular widths are moderately correlated with the LE speed. At 10 $R_\odot$, the LE speed of fast CMEs may be more strongly modulated by aerodynamic drag, whereas for slow CMEs the drag force may become more important at larger distances beyond 10 $R_\odot$ \citep{Sachdeva2015}. However, \citet{Gopalswamy2014} reported that CME widths are positively correlated with LE speed, which may arise from treating fast and slow CMEs as a single population without accounting for their distinct evolutionary behaviors. We note that the derived correlations between speeds and angular widths may be affected by uncertainties in the absolute values of the CME widths, which depend on the uncertainties in fitting GCS parameters, including the tilt angle. Uncertainties in the correlations for fast CMEs may also be reflected in the differences between the estimated face-on and edge-on angular widths in our study and those reported by \citet{Temmer2021}. A more definitive assessment will require analysis of a larger CME sample.

\subsection{Implications for CME evolution}

Our study confirms that CMEs undergo asymmetric expansion in both the radial and lateral directions, with lateral expansion (within the plane of CME propagation) persistently exceeding radial expansion along the CME propagation direction. This behavior highlights the importance of measuring expansion along different directions and demonstrates that fast and slow CMEs follow fundamentally different evolutionary paths. These differences have important implications for CME modeling, interpretation of projected observations, and space-weather forecasting, particularly when extrapolating CME properties from coronagraphic measurements to larger heliospheric distances. The asymmetric expansion, together with intrinsic inhomogeneities within the CME flux rope, can lead to variations in radial size, axis orientation, magnetic flux, and the arrival times of CME substructures at multiple in situ spacecraft (e.g., \citealt{Liu2008,Lugaz2018,Agarwal2025,Al-haddad2025}). Differences with earlier studies may also arise from sample selection effects; therefore, extending the investigation to larger samples will be essential for fully characterizing CME expansion asymmetries.

Our results further demonstrate that empirical relations assuming a constant angular width introduce systematic biases and are inadequate for modeling evolving CMEs. This underscores the need to incorporate time-dependent, de-projected geometrical parameters--such as those derived from GCS fitting--into CME kinematic models. When multi-spacecraft observations are unavailable, projected widths and speeds should therefore be interpreted with caution. Current MHD models (e.g., \citealt{Odstrcil2003,Pomoell2018,Barnard2022,Mayank2024,Owens2025}) can be used to examine the evolution of asymmetric expansion in a non-isotropic solar wind and its impact on predicting the arrival times and spatial dimensions of CME substructures at in situ spacecraft and at Earth. Incorporating asymmetric CME expansion will be crucial for improving heliospheric CME modeling.

Although our results indicate that most CMEs attain nearly constant radial and lateral expansion speeds and angular widths at larger coronagraphic heights, previous studies have shown that CME aspect ratios may continue to evolve in the interplanetary medium \citep{Agarwal2024}. Such evolution beyond the coronagraphic field of view motivates further investigation with heliospheric imagers \citep{Savani2009,Savani2011,Agarwal2024}. Coordinated observational campaigns combining EUV, white-light, and heliospheric imaging--using missions such as Aditya-L1, PROBA-3, PSP, Solar Orbiter, and PUNCH--together with time-dependent 3D modeling, will be essential for capturing the full evolution of CME expansion and propagation. We emphasize that obtaining reliable three-dimensional CME parameters is crucial, and it requires observations of CMEs from at least two distinct viewpoints \citep{Verbeke2023}. Coordinated measurements from current and upcoming missions can provide a strong pathway to improve the accuracy of CME substructures characteristics and, in turn, enhance the reliability of space-weather predictions.

\section{Conclusions}{\label{sec:conclu}}

Based on our analysis, we find that the average ratio of lateral-to-radial expansion speed at the final tracked height is $\sim$1.7 for both CME populations, indicating that asymmetric expansion remains a persistent characteristic of CME evolution within coronagraphic heights. We also find that expansion speeds derived from the widely used full cone model primarily reflect the lateral expansion of CMEs and do not reliably represent their radial expansion, highlighting an important limitation of cone-based approximations in characterizing the full 3D evolution of CMEs. Consistent with earlier studies \citep{Cremades2020}, we confirm that CMEs exhibit asymmetric expansion, with lateral expansion in the plane of propagation systematically exceeding radial expansion along the propagation direction.

Extending previous studies, which generally investigated CME expansion by treating CMEs as a single population, the present analysis separately examines the asymmetric expansion and kinematic evolution of fast and slow CMEs. This reveals that, although both populations exhibit asymmetric expansion, their evolutionary behavior is fundamentally different. Fast CMEs generally undergo rapid early expansion and acceleration at lower coronal heights and often approach a quasi-saturated state at smaller heliocentric distances, whereas slow CMEs evolve more gradually over larger heights and show a clearer coupling between expansion speeds and angular-width evolution. These results suggest that fast and slow CMEs should not be treated as a single population in statistical studies of CME properties. Overall, our findings demonstrate that time-dependent, 3D CME geometry and asymmetric expansion must be incorporated into CME models to improve the interpretation of coronagraphic observations and the reliability of space-weather predictions.

\section*{Acknowledgement}
We thank the STEREO and SOHO spacecraft team members for making their data publicly available. We also thank the anonymous referee for a careful review that improved the manuscript.

\vspace{5mm}
\facilities{STEREO(COR), SOHO(LASCO C2 and C3)}

\newpage

\appendix

\section{Description of GCS Model}\label{sec:appendix}
\begin{figure*}
\centering
\includegraphics[scale= 0.3,trim={0cm 0cm 0cm 0cm},clip]{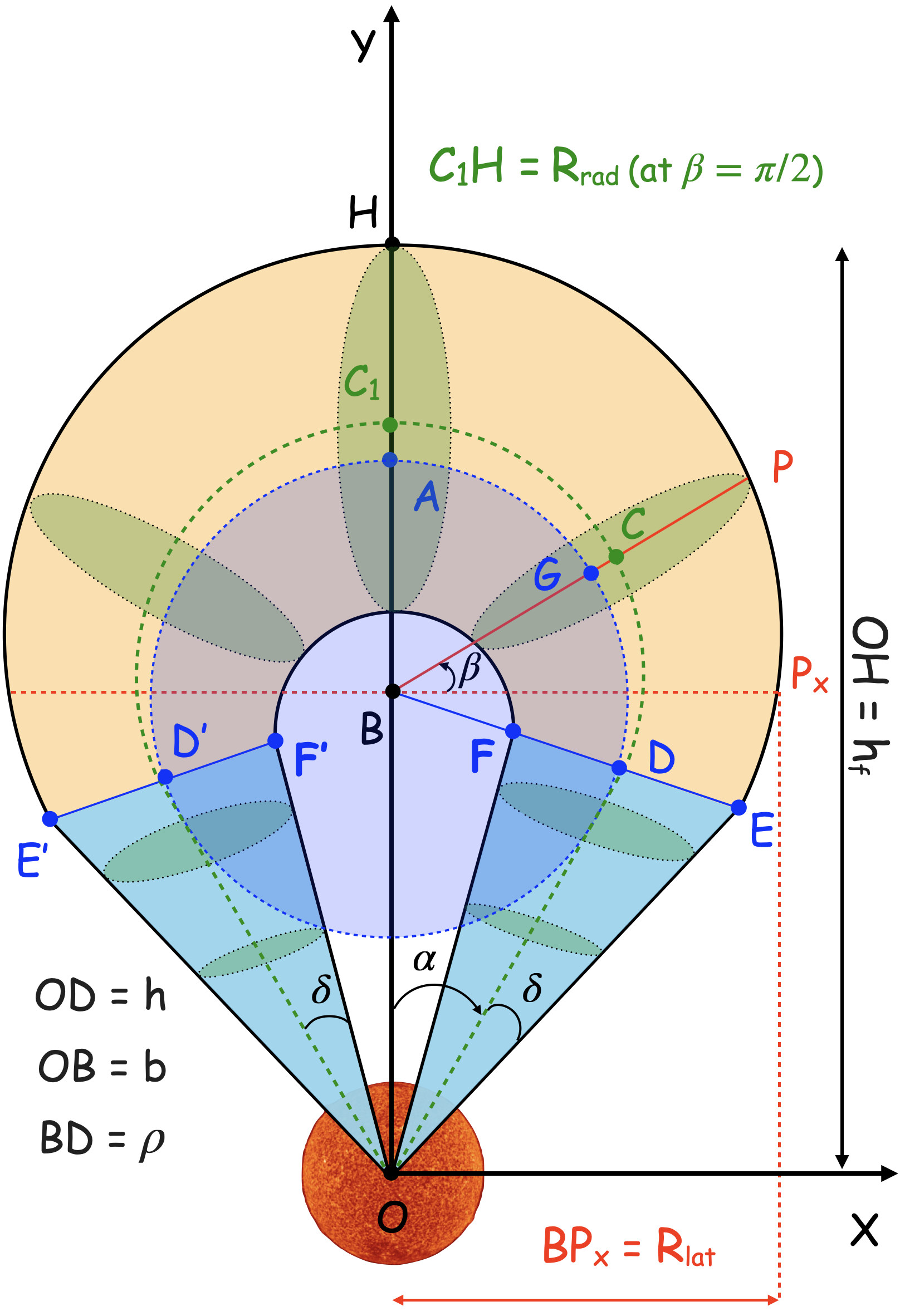}
\hspace{0.7cm} 
\includegraphics[scale= 0.3,trim={0cm 0cm 0cm 0cm},clip]{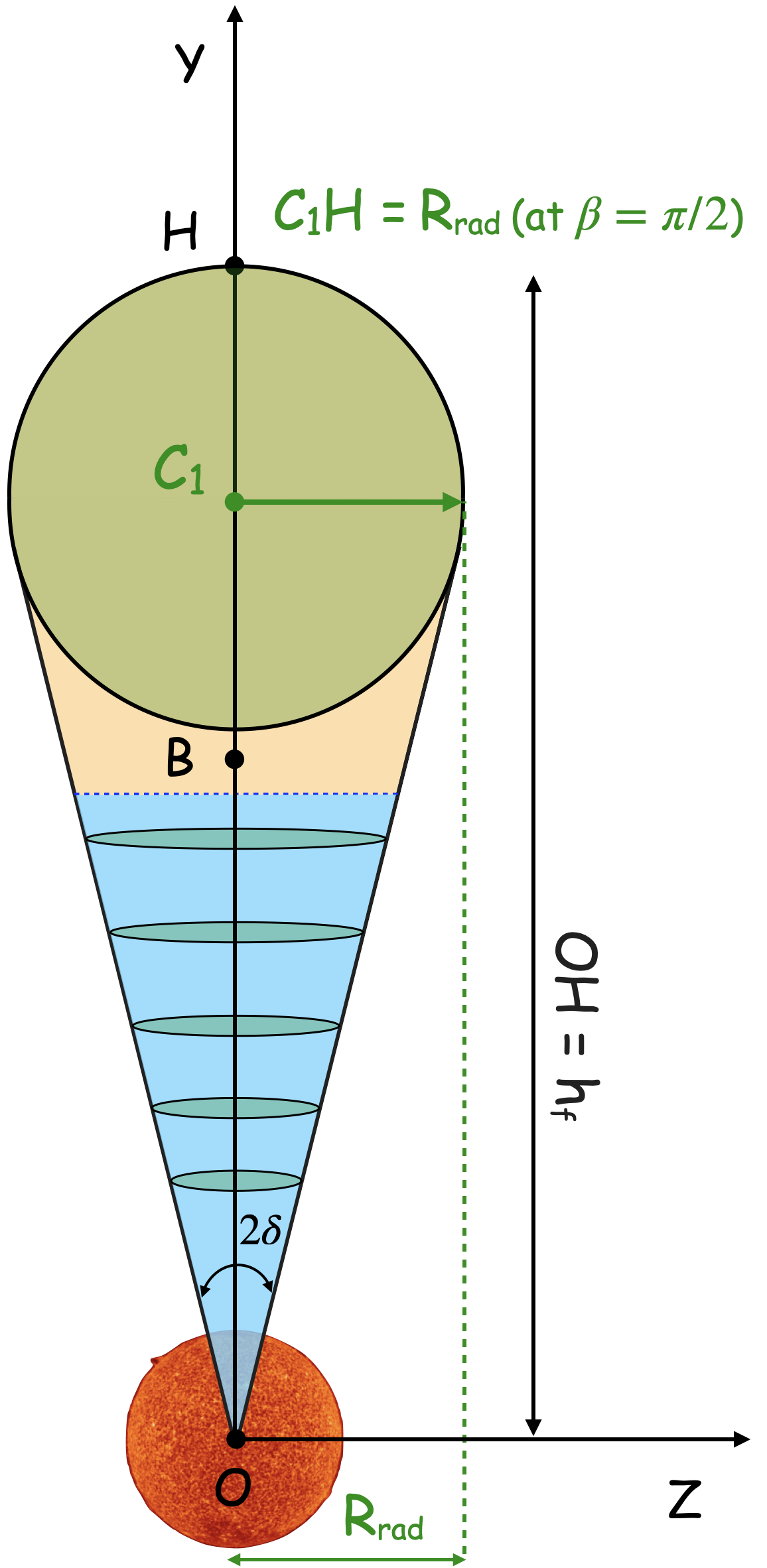}
\caption{Schematic representation of the GCS model. The left panel displays an $(O, X, Y)$  planar cut of the GCS-modeled CME geometry, illustrating its outward propagation from the Sun as viewed face-on by an observer along the $Z$ axis. This panel also shows a circle centered at the meeting point of the bases of two conical legs, and the plane of circular cross-section along the torus-shaped curved front. The right panel shows the edge-on view of the GCS-modeled CME geometry in the $(O, Y, Z)$ plane as viewed along the $X$ axis. The transparent sky blue and orange regions correspond to the two structures of the GCS model, the conical legs and the torus-shaped curved front, respectively.}
\label{fig:gcs_model}
\end{figure*}

The GCS model has been described in detail by \citet{Thernisien2011}. Here, we briefly revisit the framework and summarize the concepts most relevant to connecting the spatial and angular dimensions to the radial and lateral expansion speeds examined in the present study. In the left panel of Figure~\ref{fig:gcs_model}, the two conical legs and the torus-shaped curved front of the GCS model are shown in transparent sky blue and transparent orange, respectively. The apex of the curved front is marked by point $H$, with height $h_f=OH$ from the center of the Sun. The bases of the conical legs are given by the line segments $FE$ and $F'E'$, with their midpoints as $D$ and $D'$. The height of the conical leg is $OD = OD' = h$. The half angle between the axes of the two conical legs is  $\alpha$, and the half angle of each cone is $\delta$. The cross-section of the conical legs is circular, which lies parallel to the plane of its base, shown in transparent green. Each conical leg expands self-similarly, such that the radius of its circular cross-section satisfies $\vv{DE} = \kappa~\vv{OE}$, where $\kappa = \sin\delta$ represents the aspect ratio of the CME. The angular width observed along the toroidal direction corresponds to the face-on angular width and is equal to $2(\alpha+\delta)$.

Point $B$ marks the location on the $Y$ axis where the bases of the two conical legs meet, and the distance from the origin is given by $OB = b = h/ \cos\alpha$. The circle of radius $BD$ ($\rho = h\tan\alpha$), shown in transparent blue in the $(O, X, Y)$ plane, serves as the generating line for the circular cross-sections (transparent green) of the torus-shaped (transparent orange) curved front, and the radius of the cross-sections increases with height. Each circular cross-section of the torus-shaped front lies in the plane $(B, \vv{BG}, Z)$, where $G$ is a point on the transparent blue circle, and the angle between the $X$ axis and the vector $\vv{BG}$ is $\beta$. The curved front follows the self-similarity constraint expressed by $\vv{GP} = \kappa~\vv{OP}$, where $P$ is a point on the croissant shell that is not constrained to lie in the $(O, X, Y)$ plane.

\begin{table*}
    \centering
    \small
    {\renewcommand{\arraystretch}{1.8}
    \begin{tabular}{lcc}
    \hline
    {Parameters} & {Expression} & \\
    \hline
    {{{Distance of the CME center ($OC_1 = h_c$)}}} & {$b +X_0(\beta = \pi/2) = \dfrac{b+\rho}{1-\kappa^2}$} & (a)\\
    
    {{Height of the CME apex ($OH$)}} & {{$h_f = \dfrac{b+\rho}{1-\kappa} = OC_1(1+\kappa)$}} & (b)\\
    
    {{Cross-section radius of the curved front}} & {{$R(\beta = \pi/2) = \dfrac{\kappa(b + \rho)}{1-\kappa^2}$}} & (c)\\
    
    {{Face-on angular width (along the toroidal direction)}} & {{$\omega_{FO} = 2(\alpha + \delta)$}} & (d)\\
    
    {{Edge-on angular width (along the poloidal direction)}} & {{$\omega_{EO} = 2\delta$}} & (e)\\
    
    {{Aspect ratio}} & {{$\kappa = \dfrac{R_{rad}}{OC_1}$}} & (f)\\

    {{$x$ coordinate of $\vv{BP}~(BP_x)$}} & {{$\dfrac{h[\kappa + sin\alpha(1 + \kappa) + \kappa^2sin\beta]cos\beta}{(1 + \kappa)(1 + sin\alpha)}$}} & (g) \\
    
    {{$R_{lat}$}} & {{max of $BP_x$ at $\beta$ where $\dfrac{d(BP_{x})}{d\beta} = 0$}} & (h)\\
    
    {{$R_{rad}$}} & {{$(\dfrac{\kappa}{1+\kappa})h_f$}} & (i)\\
    
    {{Lateral dimension}} & {{$2R_{lat}$}} & (j)\\
    
    {{Radial dimension}} & {{$2R_{rad}$}} & (k)\\
    
    {{Radial expansion speed}} & {{$V_{exp\_rad} = \dfrac{d(R_{rad})}{dt}$}} & (l)\\
    
    {{Lateral expansion speed}} & {{$V_{exp\_lat} = \dfrac{d(R_{lat})}{dt}$}} & (m)\\
    \hline 
    \end{tabular}}
    \caption{The table lists the formulae for calculating the GCS model parameters.}
    \label{tab:gcs_parameter}
\end{table*}

The centers of the circular cross-sections of both the torus-shaped curved front and the conical legs lie on the flux-rope axis, which is shown by the green dotted curved line. The distance between point $B$ and the center of the circular cross-section $C$ (in the $(O, X, Y)$ plane) is denoted by $X_0$. The radius of the circular cross sections ($R$) of the torus-shaped curved front and the distance $X_0$ depend on the angle $\beta$ and on the parameters $\rho$, $\kappa$, and $b$. The expressions for $X_0$ and $R$ are given by Equations 18 and 19 of \citet{Thernisien2011}, and are also shown below as they are crucial for estimating the one radial and two lateral dimensions of the GCS modeled CME structure.

\begin{equation*}
X_{0} = \frac{\rho + b\kappa^2 sin\beta}{1 - \kappa^2}
\end{equation*}

\begin{equation*}
R^2 = X_0^2 + \frac{b^2 \kappa^2 - \rho^2}{1 - \kappa^2}
\end{equation*}

The radius of the circular cross-section attains its maximum at $\beta = \pi/2$. This value corresponds to half of the lateral dimension of the GCS structure in the plane perpendicular to the CME propagation direction. Owing to the adopted GCS geometry, this lateral dimension (in the plane perpendicular to the CME propagation) is equal to the radial dimension (in the plane of CME propagation) of the CME flux rope. However, the lateral dimension within the plane of CME propagation differs from the radial dimension. As illustrated in the figure, the maximum radius of the toroidal cross-section, denoted by $C_1H$, represents half of the radial dimension $R_{rad}$, where $C_1$ is the center of the circular cross-section containing the apex of the curved front. When point $P$ is bound to lie in the $(O, X, Y)$ plane, the components of the vector $\vv{BP}$ are given by $([X_0 + R]\cos\beta, [X_0 + R]\sin\beta, 0)$. The $x$-component of $\vv{BP}$ is listed as Equation (g) of Table~\ref{tab:gcs_parameter}, and the maximum value of it (denoted as $BP_x$ in the middle panel) corresponds to half of the lateral (face-on) dimension of the torus-shaped curved front within the plane of CME propagation. The value of $\beta$ at which $BP_{x}$ is maximum and becomes $R_{lat}$ can be estimated by setting the derivative of $BP_{x}$ with respect to $\beta$ equal to zero, as expressed below:

\begin{align}\label{equ:2}
\frac{d(BP_{x})}{d\beta} = \kappa^2 (\cos^2\beta - \sin^2\beta) - \kappa \sin\beta(1 + \sin\alpha) \\
- \sin\alpha \sin\beta = 0 \nonumber
\end{align}

The right panel of the figure represents the edge-on view of the GCS structure as seen by an observer located along the $X$ axis. In this figure, the angular width observed corresponds to the edge-on angular width and is equal to $2\delta$. The edge-on angular width is termed as angular width along the poloidal (in planes perpendicular to the GCS model axis) direction. The maximum radius of the toroidal cross-section, $R_{rad}$ ($C_1H$), is derived using Equations (b) and (f), and is listed as Equation (i) in Table~\ref{tab:gcs_parameter}.

\end{document}